\documentclass[a4paper,reqno]{amsart}

\usepackage[T1]{fontenc}
\usepackage[english]{babel}	
\usepackage{amssymb}
\usepackage[protrusion=true,expansion=true]{microtype}	
\usepackage{amsmath,amsfonts,amsthm} 
\usepackage[pdftex]{graphicx}	
\usepackage{url}
\usepackage[title,titletoc,page]{appendix} 
\usepackage{listings} 
\usepackage{color}
\usepackage{comment}
\usepackage{caption}
\usepackage{subcaption}
\usepackage{algorithm}
\usepackage[foot]{amsaddr}

\usepackage{todonotes}

\usepackage{adjustbox}

\usepackage{tikz}
\usetikzlibrary{matrix,positioning,decorations.pathreplacing,calc}
\usetikzlibrary{
decorations.text,%
decorations.markings,%
shadows}

\usepackage{mathtools}

\DeclareMathOperator{\iL}{{\mathsf{L}}}
\DeclareMathOperator{\iR}{{\mathsf{R}}}
\DeclareMathOperator{\iLR}{{\mathsf{L},\mathsf{R}}}

\newcommand{\pri}{^\prime}
\newcommand{\prii}{^{\prime\prime}}

\DeclareMathOperator{\dd}{d\!}
\usepackage{bm}

\usepackage{fancyhdr}

\usepackage[%
pdfauthor={Habib Ammari, Silvio Barandun, Ping Liu},%
pdftitle={Perturbed Block Toeplitz matrices and the non-Hermitian skin effect in dimer systems of subwavelength resonators},
pdfsubject={Perturbed Block Toeplitz matrices and the non-Hermitian skin effect in dimer systems of subwavelength resonators},
pdfkeywords={Block Toeplitz matrix, perturbation, eigenvalue, eigenvector, 
Non-Hermitian skin effect, subwavelength physics},
]{hyperref}

\usepackage[sort,nocompress]{cite}

\numberwithin{equation}{section}		
\numberwithin{figure}{section}			
\numberwithin{table}{section}				
\usepackage[noabbrev,capitalize]{cleveref}
\newtheorem{thm}{Theorem}[section]
\newtheorem{defi}[thm]{Definition}

\newtheorem{remark}[thm]{Remark}
\newtheorem{prop}[thm]{Proposition}

\newcommand{\babs}[1]{\left|{#1}\right|}

\newcommand{\vect}[1]{\boldsymbol{\mathbf{#1}}}

\newcommand{\R}{\mathbb{R}}

\DeclareMathOperator{\diag}{diag}

\title[Non-Hermitian skin effect in dimer systems]{Perturbed Block Toeplitz matrices and the non-Hermitian skin effect in dimer systems of subwavelength resonators}

\author[H. Ammari]{Habib Ammari\textsuperscript{1}}
\address{\textsuperscript{1}ETH Zürich, Department of Mathematics, Rämistrasse 101, 8092 Zürich, Switzerland.}
\email{habib.ammari@math.ethz.ch}
\author[S. Barandun]{Silvio Barandun\textsuperscript{1}}
 \email{silvio.barandun@sam.math.ethz.ch}
\author[P. Liu]{Ping Liu\textsuperscript{1}}
\email{ping.liu@sam.math.ethz.ch}

\begin{document}

\maketitle

\begin{abstract}
The aim of this paper is fourfold: (i) to obtain explicit formulas for the eigenpairs of perturbed tridiagonal block Toeplitz matrices; (ii) to make use of such formulas in order to provide  a mathematical justification of the non-Hermitian skin effect in dimer systems by proving the condensation of the system's bulk eigenmodes at one of the edges of the system; (iii) to show the topological origin of the non-Hermitian skin effect for dimer systems and (iv) to prove localisation  of the interface modes between two dimer structures with non-Hermitian gauge potentials of opposite signs based on new estimates of the decay of the entries of the eigenvectors of block matrices with mirrored blocks.
\end{abstract}

\date{}

\bigskip

\par

\bigskip

\\

\section{Introduction}

 The ultimate goal of subwavelength  physics is to manipulate waves at subwavelength scales in a robust way \cite{lemoult.fink.ea2011Acoustic,lemoult.kaina.ea2016Soda,yves.fleury.ea2017Crystalline,sheng,phononic1,ammari.davies.ea2021Functional}. 
 Subwavelength resonators are the building blocks of the resonant structures used in subwavelength physics. Many spectacular phenomena  in subwavelength  physics  have been recently demonstrated and mathematically studied \cite{ammari.davies.ea2021Functional,ammari.davies.ea2022Exceptional,ammari.davies.ea2020Topologically,florian,ammari.barandun.ea2023Edge,craster.davies2022Asymptotic}. The non-Hermitian skin effect is one of the most intriguing ones \cite{yokomizo.yoda.ea2022NonHermitian,zhang.zhang.ea2022review}. Due to a complex gauge potential inside the resonators,  for a system of finitely many resonators the eigenmodes decay exponentially and condensate at one of the edges of the structure.  In \cite{ammari2023mathematical},  a mathematical theory of the non-Hermitian skin effect arising in subwavelength physics in one dimension has been derived from first principles. Through a gauge capacitance matrix formulation, explicit asymptotic expressions for the subwavelength eigenfrequencies and eigenmodes of systems of a single repeating resonator have been obtained. This allowed the authors to characterise the system's fundamental behaviours and reveal the mechanisms behind them. In particular, the exponential decay of eigenmodes (the so-called non-Hermitian skin effect)  was shown to be induced by the Fredholm index of an associated 
  (tridiagonal)  Toeplitz operator. The explicit theory developed in \cite{ammari2023mathematical} was only possible because of the simple structure of the gauge capacitance matrix and the rich literature on (tridiagonal) Toeplitz matrices and perturbations thereof \cite{noschese.pasquini.ea2013Tridiagonal,yueh.cheng2008Explicit}. The theory of systems with periodically repeated cells of $K$ resonators ($K \geq 2$) remains incomplete as no similar results to those used in \cite{ammari2023mathematical} have been known for block-Toeplitz matrices. However, some  numerical illustrations of the non-Hermitian skin effect in systems of finitely many dimers with imaginary gauge potentials are presented. These numerical results clearly show strongly localised eigenmodes at one edge of the dimer system and suggest that the non-Hermitian skin effect holds for systems with multiple resonators in the unit cell. Nevertheless, compared to the single resonator case, the physics for dimer systems is much richer as the system eigenvalues are grouped into two families corresponding to eigenmodes with two different physical natures (monopole and dipole behaviors). Consequently, the mathematical analysis of dimer systems is much harder.  

In this paper, we obtain for the first time  explicit formulas for the eigenpairs of perturbed tridiagonal block Toeplitz matrices, which have their own interest and may found applications in other fields such as quantum mechanics and condensed matter theory. 
Applying these formulas in the field of subwavelength physics, we provide  a mathematical justification of the non-Hermitian skin effect in dimer systems by proving the condensation of the eigenvectors of the associated gauge capacitance matrix. Moreover, 
we show the topological origin of the non-Hermitian skin effect for dimer systems.  In contrast with the single resonator case, 
 the determinant of the symbol of the $2$-Toeplitz operator associated with the semi-infinite structure has a zero on the unit circle, and therefore its winding is not equal to the Fredholm index of the operator.  Nevertheless, since the system eigenvalues are grouped into two families corresponding to eigenmodes with monopole and dipole behaviors, we can show that each group corresponds to negative winding of one of the eigenvalues of the symbol of the $2$-Toeplitz operator. On the other hand, we consider interface modes between two structures where the sign of  the complex gauge potential changes and prove that all but few eigenmodes are localised at the interface between the two structures. 

The paper presents a number of original results and findings: (i) a general strategy for deriving formulas for the eigenvalues and eigenvectors of tridiagonal $2$-Toeplitz matrices with perturbations on the diagonal corners; (ii) an estimate of the decay of the entries of the eigenvectors of block matrices with mirrored blocks; and (iii) mathematical foundations of the non-Hermitian skin effect in dimer systems, its topological origin and non-Hermitian interface modes between opposing signs of the gauge potentials. 

The paper is organised as follows. In Section \ref{section:eigenvalueanalysis1}, we first recall some known results on Chebyshev polynomials and then characterize the eigenvalues of tridiagonal $2$-Toeplitz matrices with perturbations on the diagonal corners. Section \ref{sect3} is dedicated to the construction of the eigenvectors of tridiagonal 2-Toeplitz matrices with perturbations on the diagonal corners. In Section \ref{sect4}, we prove the condensation of the eigenvectors of  perturbed $2$-Topelitz matrices and block matrices with mirrored  $2$-Toeplitz matrices.   
In Section \ref{sect5}, we formulate the physical model for dimer systems of subwavelength resonators with a complex gauge potential inside only the resonators. Without exciting the 
 structure's subwavelength resonances, the effect of the complex gauge potential would be negligible. We also   show  
how to apply the general results obtained in the previous sections to prove the non-Hermitian skin effect for dimer systems and the eigenmode condensation at the interface between two structures with opposite signs of gauge potentials. Furthermore, we provide the topological origin of the non-Hermitian skin effect for dimer systems.
In  Section \ref{sect6}, we draw some conclusions and state some open problems and extensions to our present work.

\section{Eigenvalues of tridiagonal $2$-Toeplitz matrices with perturbations on the diagonal corners}\label{section:eigenvalueanalysis1}

In this section, we first present some well-known results on Chebyshev polynomials and then characterize the eigenvalues of tridiagonal $2$-Toeplitz matrices with perturbations on the diagonal corners. 

\subsection{Tridiagonal $2$-Toeplitz matrices with perturbations on the diagonal corners}
Let  $A_{2m+1}^{(a,b)}$ be the tridiagonal $2$-Toeplitz matrix of order $2m+1$ with perturbations $(a, b)$ in the diagonal corners, that is, 
\begin{equation}\label{equ:oddmatrixtwoperturb1}
A_{2m+1}^{(a, b)}=\left(\begin{array}{ccccccc}
    \alpha_{1}+a & \beta_{1} & & & & &\\
    \gamma_{1} & \alpha_{2} & \beta_{2} & & & &\\
    & \gamma_{2} & \alpha_{1} & \beta_{1} & & &\\
    & & \gamma_{1} & \alpha_{2} & \ddots & &\\
    & & & \ddots & \ddots&\ddots&\\
    & & &  &\gamma_{1} & \alpha_2 & \beta_2\\
    & & &  & & \gamma_{2} & \alpha_{1}+ b\\
\end{array}\right).
\end{equation}
Here, $\beta_i, \gamma_i, \alpha_i$, $i=1,2$ and $a,b$ are in $\mathbb{R}$. 
Let  $A_{2m}^{(a,b)}$ be the tridiagonal $2$-Toeplitz matrix of order $2m$ with perturbations $(a, b)$ in the diagonal corners, that is, 
\begin{equation}\label{equ:evenmatrixtwoperturb1}
A_{2m}^{(a,b)}=\left(\begin{array}{ccccccc}
    \alpha_{1}+a & \beta_{1} & & & & &\\
    \gamma_{1} & \alpha_{2} & \beta_{2} & & & &\\
    & \gamma_{2} & \alpha_{1} & \beta_{1} & & &\\
    & & \gamma_{1} & \alpha_{2} & \ddots & &\\
    & & & \ddots & \ddots&\ddots&\\
    & & &  &\gamma_{2} & \alpha_1 & \beta_1\\
    & & &  & & \gamma_{1} & \alpha_{2}+ b\\
\end{array}\right).
\end{equation}

\begin{remark}
We assume throughout that the off-diagonal elements in (\ref{equ:oddmatrixtwoperturb1}) and (\ref{equ:evenmatrixtwoperturb1}) are nonzero and satisfy the following condition: $$\gamma_i\beta_i >0, \quad i=1,2.$$
\end{remark}


\subsection{Chebyshev polynomials}
 The Chebyshev polynomials are two sequences of polynomials related to the cosine and sine functions, denoted respectively by $T_{n}(x)$ and $U_{n}(x)$. In particular, the Chebyshev polynomials of the first kind are obtained from the recurrence relation
\begin{align*}
T_0(x) & =1, \\
T_1(x) & =x, \\
T_{n+1}(x) & =2 x T_n(x)-T_{n-1}(x) ,
\end{align*}
and the Chebyshev polynomials of the second kind are obtained from the recurrence relation
\begin{align*}
U_0(x) & =1, \\
U_1(x) & =2x, \\
U_{n+1}(x) & =2 x U_n(x)-U_{n-1}(x) .
\end{align*}
The roots of $T_n(x)$ are
\[
x_k=\cos \left(\frac{\pi(k+1 / 2)}{n}\right), \quad k=0, \ldots, n-1,
\]
and the roots of $U_{n}(x)$ are 
\begin{equation}\label{equ:rootsofchebyshevpolynomial1}
x_k = \cos\left(\frac{k\pi}{n+1}\right), \quad k  =1, \cdots, n.
\end{equation}
It is also well-known that for $-1 \leq x \leq 1$ and $k=0,1, \ldots$, we have the upper bounds
\begin{equation}\label{equ:boundofchebyshev1}
\left|T_n(x)\right| \leq\left|T_n(1)\right|=1, \quad \left|U_k(x)\right| \leq\left|U_k(1)\right|=k+1.
\end{equation}

\subsection{Eigenvalues of $A_{2m+1}^{(a,b)}$ and $A_{2m}^{(a,b)}$}
In this subsection, we present a detailed characterisation of the eigenvalues of $A_{2m+1}^{(a,b)}$ and $A_{2m}^{(a,b)}$.	
	
Let us first define the polynomials
$$
\pi_2(x)=\left(x-\alpha_1\right)\left(x-\alpha_{2}\right)
$$
and
\begin{equation}\label{equ:defiofpkstar1}
P_k^*(x)=\left(\sqrt{\gamma_{1}\beta_1 \gamma_2\beta_{2}}\right)^k U_k\left(\frac{x-\gamma_{1}\beta_1-\gamma_2\beta_2}{2 \sqrt{\gamma_{1}\beta_{1} \gamma_{2}\beta_2}}\right), 
\end{equation}
where $U_k$ is the Chebyshev polynomial of the second kind. It is well known (see \cite{gover1994eigenproblem, da2001explicit, marcellan1999orthogonal} ) that the characteristic polynomials of the $2$-Toeplitz matrices $A_{2m+1}^{(0,0)}, A_{2m}^{(0,0)}$ are respectively 
\begin{equation}\label{equ:eigenpolynomial1}
Q_{2 k+1}(x)=\left(x-\alpha_{1}\right) P_k^*\left(\pi_2(x)\right)
\end{equation}
and
\begin{equation}\label{equ:eigenpolynomial2}
Q_{2 k}(x)=P_k^*\left(\pi_2(x)\right)+\gamma_{2}\beta_{2} P_{k-1}^*\left(\pi_2(x)\right).
\end{equation}	
It is also shown in \cite{da2007characteristic} that the characteristic polynomials of $A_{2m+1}^{(a, b)}, A_{2m}^{(a, b)}$ are respectively 
\begin{equation}\label{equ:eigenpolynomial3}
\begin{aligned}
P_{2m+1}(x) =& \left(x-\alpha_{1}-a-b\right) P_m^*\left(\pi_2(x)\right)\\
& +\left( ab\left(x-\alpha_{2}\right)  -a \gamma_{1}\beta_{1}-b \gamma_{2}\beta_{2}\right) P_{m-1}^*\left(\pi_2(x)\right)
\end{aligned}
\end{equation}
and 
\begin{equation}\label{equ:eigenpolynomial4}
\begin{aligned}
 P_{2m}(x)
 =&P_m^*\left(\pi_2(x)\right)+\left(a\left(\alpha_2-x\right)+b\left(\alpha_1-x\right)+a +b+\gamma_{2}\beta_{2}\right) P_{m-1}^*\left(\pi_2(x)\right)\\
& +a b \gamma_1\beta_{1} P_{m-2}^*\left(\pi_2(x)\right).
\end{aligned}
\end{equation}
For more results on the eigenproblem of tridiagonal $K$-Toeplitz matrices, we refer the readers to \cite{da2007characteristic} and the references therein.

To help to demonstrate the non-Hermitian skin effect, we first present a detailed characterisation of the eigenvalues of $A_{2m+1}^{(a, b)}$ and $A_{2m}^{(a, b)}$. In particular, we remark that one may have better results through a delicate analysis on the roots of (\ref{equ:eigenpolynomial3}) and (\ref{equ:eigenpolynomial4}), but here we choose a simple method using the Cauchy interlacing theorem.

\begin{thm}\label{thm:eigenvaluethm1}
 Let $m$ be large enough. The eigenvalues $\lambda_r$'s of $A_{2m+1}^{(a, b)}$ are all real numbers. Except for at most $11$ eigenvalues, we can reindex the $\lambda_r$'s to have
 \[
 \lambda_{3}^l < \lambda_{4}^l<\cdots< \lambda_{m-3}^l\leq \min\{\alpha_1, \alpha_2\}\leq  \max\{\alpha_1, \alpha_2\}\leq \lambda_{m-3}^r< \cdots< \lambda_{4}^{r}< \lambda_{3}^{r}.
\]
In particular, for $k =3,\cdots, m-3$, 
\begin{equation}\label{equ:eigenvalue1}
\cos\left(\frac{k\pi}{m}\right)\leq \min\left\{y\left(\lambda_k^l\right), y\left(\lambda_k^r\right)\right\}\leq \max\left\{y\left(\lambda_k^l\right), y\left(\lambda_k^r\right)\right\} \leq \cos\left(\frac{(k-2)\pi}{m}\right), 
\end{equation}
with 
\begin{equation}\label{equ:normalizedfunction1}
y(x)=\frac{(x-\alpha_1)(x-\alpha_2)-\gamma_1\beta_1-\gamma_2\beta_2}{2\sqrt{\gamma_1\beta_1\gamma_2\beta_2}}.
 \end{equation}
\end{thm}

 \begin{proof} The proof is divided into two steps. 

 \noindent
 \textbf{Step 1.} Note that $A_{2m+1}^{(a,b)}$ has the same eigenvalues as the Hermitian matrix
 \begin{equation}\label{equ:hermitianmatrix1}
	H:=\left(\begin{array}{ccccccc}
		\alpha_{1}+a & \sqrt{\gamma_1\beta_{1}} & & & & &\\
		\sqrt{\gamma_{1}\beta_1} & \alpha_{2} & \sqrt{\gamma_2\beta_{2}} & & & &\\
		& \sqrt{\gamma_{2}\beta_2} & \alpha_{1} & \sqrt{\gamma_1\beta_{1}} & & &\\
		& & \sqrt{\gamma_{1}\beta_1} & \alpha_{2} & \ddots & &\\
		& & & \ddots & \ddots&\ddots&\\
		& & &  &\sqrt{\gamma_{1}\beta_1} & \alpha_2 & \sqrt{\gamma_2\beta_2}\\
		& & &  & & \sqrt{\gamma_{2}\beta_2} & \alpha_{1}+ b\\
	\end{array}\right)
 \end{equation}
 which can seen from computing $\babs{xI-A_{2m+1}^{(a,b)}} = \babs{xI-H}$ using Laplace expansion on the last row. 
 Thus the eigenvalues of $A_{2m+1}^{(a,b)}$ are real numbers.  To 
 demonstrate (\ref{equ:eigenvalue1}), we first 
 analyse the eigenvalues of $A_{2m+1}^{(0,0)}$, i.e., the case when $a=b=0$. Note 
 that by (\ref{equ:eigenpolynomial1}) the eigenvalues $\{\lambda_r\}$ of $A_{2m+1}^{(0,0)}$ are the roots 
 of the polynomial
 \[
Q_{2m+1}(x) = \left(x-\alpha_{1}\right) P_m^*\left(\pi_2(x)\right).
 \]
By the definition of $P_{k}^*(x)$ in (\ref{equ:defiofpkstar1}), to find the roots of $Q_{2m+1}(x)$ (except the trivial root $x=\alpha_1$), we only need to find the solutions of
\begin{align}\label{equ:proofeigenvalueequ-1}
 U_{m}(y)=0,
\end{align}
where $y$ is defined by (\ref{equ:normalizedfunction1}). By (\ref{equ:rootsofchebyshevpolynomial1}), the solutions are given by
\[
y_k = \cos\left(\frac{k\pi}{m+1}\right), \ k=1,2,\cdots, m.
\]
From (\ref{equ:normalizedfunction1}), it is not hard to see that the $\lambda_r$'s corresponding to $y_k = \cos\left(\frac{k\pi}{m+1}\right)$ should belong to $\left(-\infty, \min\{\alpha_1, \alpha_2\}\right]$ or $\left[\max\{\alpha_1, \alpha_2\},+\infty\right)$. Therefore, the $\lambda_r$'s can be reindexed in order to have 
\[
 \lambda_{1}^l < \lambda_{2}^l<\cdots< \lambda_{m}^l\leq \min\{\alpha_1, \alpha_2\}\leq  \max\{\alpha_1, \alpha_2\}\leq \lambda_{m}^r< \cdots< \lambda_{2}^{r}< \lambda_{1}^{r}
\]
with 
\begin{equation*}
y\left(\lambda_k^l\right) = y\left(\lambda_k^r\right)= \cos\left(\frac{k\pi}{m+1}\right), \quad k =1,2,\cdots, m.
\end{equation*}

\noindent \textbf{Step 2.} We now turn to the case when $(a, b)\neq (0,0)$. Consider $A_{2m+1}^{(a,b)}$'s principal submatrices $A_{2m}^{(a,0)}$ and 
\[
D_{2m-1}^{(0,0)}= \left(\begin{array}{ccccc}
		 \alpha_{2} & \beta_{2} & & &\\
		 \gamma_{2} & \alpha_{1} & \beta_{1} & &\\
		 & \gamma_{1} & \alpha_{2} & \ddots &\\
		 & & \ddots & \ddots&\beta_{2}\\
		 & &  &\gamma_{2} & \alpha_1 
	\end{array}\right).
\]

Denote the eigenvalues of $A_{2m}^{(a,0)}$ by $t_1,\cdots, t_{2m}$, assuming that they are distributed in decreasing order and the eigenvalues of $D_{2m-1}^{(0,0)}$ by $g_1,\cdots, g_{2m-1}$, assuming  that they are distributed in decreasing order. Since the eigenvalues of $A_{2m}^{(a,0)}$ and $D_{2m-1}^{(0,0)}$ are the same as those of some Hermitian matrices like (\ref{equ:hermitianmatrix1}), by the Cauchy interlacing theorem, we thus obtain that
\begin{equation}\label{equ:proofeigenvalueequ1}
t_{2m} \leq g_{2m-1} \leq t_{2m-1} \leq g_{2m-2} \leq \cdots \leq t_2 \leq g_1 \leq t_1.
\end{equation}
Applying the same arguments as those in \textbf{Step 1} to the eigenvalues $\{g_r\}$ of $D_{2m-1}^{(0,0)}$, we have that the $g_r$'s can be reindexed to have
\[
g_{1}^l < g_{2}^l < \cdots< g_{m-1}^l\leq \min\{\alpha_1, \alpha_2\}\leq  \max\{\alpha_1, \alpha_2\}\leq g_{m-1}^r< \cdots< g_{2}^{r}< g_{1}^{r}
\]
with
\begin{equation}\label{equ:proofeigenvalueequ4}
y\left(g_k^l\right) = y\left(g_k^r\right)= \cos\left(\frac{k\pi}{m}\right), \quad k =1,\cdots, m-1.
\end{equation}
By \eqref{equ:proofeigenvalueequ1}, except for at most $6$ $t_r$'s, the eigenvalues of $A_{2m}^{(a, 0)}$ can be reindexed to have
\[
 t_{2}^l < t_{3}^l<\cdots< t_{m-2}^l\leq \min\{\alpha_1, \alpha_2\}\leq  \max\{\alpha_1, \alpha_2\}\leq t_{m-2}^r< \cdots< t_{3}^{r}< t_{2}^{r}.
\]
In particular, since (\ref{equ:normalizedfunction1}) is decreasing on the left of $\frac{\alpha_1+\alpha_2}{2}$ and increasing on the right, by (\ref{equ:proofeigenvalueequ4}) we have for $k=2,3,\cdots, m-2$,
\[
\cos\left(\frac{k\pi}{m}\right)\leq \min\left\{y\left(t_k^l\right), y\left(t_k^r\right)\right\}\leq \max\left\{y\left(t_k^l\right), y\left(t_k^r\right)\right\}\leq  \cos\left(\frac{(k-1)\pi}{m}\right). 
\]
Similarly, as $A_{2m}^{(a,0)}$ is a principal submatrix of $A_{2m+1}^{(a,b)}$, by the Cauchy interlacing theorem, we have that, except for at most $11$ eigenvalues, we can arrange the eigenvalues in such a way that 
\[
 \lambda_{3}^l < \lambda_{4}^l<\cdots< \lambda_{m-3}^l\leq \min\{\alpha_1, \alpha_2\}\leq  \max\{\alpha_1, \alpha_2\}\leq \lambda_{m-3}^r< \cdots< \lambda_{4}^{r}< \lambda_{3}^{r}.
\]
In particular, for $k =3,\cdots, m-3$, 
\begin{equation*}
\cos\left(\frac{k\pi}{m}\right)\leq \min\left\{y\left(\lambda_k^l\right), y\left(\lambda_k^r\right)\right\}\leq \max\left\{y\left(\lambda_k^l\right), y\left(\lambda_k^r\right)\right\} \leq \cos\left(\frac{(k-2)\pi}{m}\right).
\end{equation*}
This completes the proof. 

\end{proof}

\medskip
We now prove the following result. 
\begin{thm}\label{thm:eigenvaluethm2}
Let $m$ be large enough. The eigenvalues $\{\lambda_r\}$ of $A_{2m}^{(a, b)}$ are all real numbers. Except for at most $12$ eigenvalues, we can reindex the $\lambda_r$'s to have
 \[
 \lambda_{3}^l < \lambda_{4}^l<\cdots< \lambda_{m-4}^l\leq \min\{\alpha_1, \alpha_2\}\leq  \max\{\alpha_1, \alpha_2\}\leq \lambda_{m-4}^r< \cdots< \lambda_{4}^{r}< \lambda_{3}^{r}.
\]
In particular, for $k =3,\cdots, m-4$, 
\begin{equation}\label{equ:eigenvalue2}
\cos\left(\frac{(k+1)\pi}{m}\right)\leq \min\left\{y\left(\lambda_k^l\right), y\left(\lambda_k^r\right)\right\}\leq \max\left\{y\left(\lambda_k^l\right), y\left(\lambda_k^r\right)\right\} \leq \cos\left(\frac{(k-2)\pi}{m}\right)
\end{equation}
with $y(x)$ being defined by (\ref{equ:normalizedfunction1}). 
\end{thm}
\begin{proof}
The proof is divided into two steps. 

\noindent \textbf{Step 1.} Similar to the case of $A_{2m+1}^{(a,b)}$, the eigenvalues of $A_{2m}^{(a,b)}$ are all real numbers. To demonstrate (\ref{equ:eigenvalue2}), we first 
 analyse the eigenvalues of $A_{2m}^{(0,0)}$. Note 
 that by (\ref{equ:eigenpolynomial2})  the eigenvalues $\{\lambda_r\}$ of $A_{2m}^{(0,0)}$ are the roots of
 \[
Q_{2m}(x) = P_m^*\left(\pi_2(x)\right)+\gamma_{2}\beta_{2} P_{m-1}^*\left(\pi_2(x)\right).
 \]
Similarly, in order to find the roots of $Q_{2m}(x)$, we only need to find the solutions $\{y_k\}$ to
\begin{equation}\label{equ:proofeigenvalueequ0}
U_{m}(y)+\sqrt{\frac{\gamma_2\beta_2}{\gamma_1\beta_1}}U_{m-1}(y)=0.
\end{equation}
Define 
\[
f(y)= U_{m}(y)+\sqrt{\frac{\gamma_2\beta_2}{\gamma_1\beta_1}}U_{m-1}(y).
\]
Without loss of generality, we suppose that $m$ is an odd number. Note that the roots of $U_{m-1}(y)$ are $\cos \frac{k\pi}{m}, \ k=1,2,\cdots, m-1$ and 
\begin{equation}\label{equ:proofeigenvalueequ-2}
\begin{array}{ll}
U_{m-1}(y)> 0 &\text{for $y \in \left(\cos \left(\frac{(2k+1)\pi}{m}\right),\ \cos \left(\frac{2k\pi}{m}\right)\right), \ k=0, \cdots, \frac{m-1}{2},$} \\
U_{m-1}(y)< 0 &\text{for $y \in \left(\cos \left(\frac{(2k+2)\pi}{m}\right),\ \cos \left(\frac{(2k+1)\pi}{m}\right) \right),  \ k=0, \cdots, \frac{m-3}{2}.$}
\end{array}
\end{equation}
Note also that
\[
\frac{k-1}{m} < \frac{k}{m+1}< \frac{k}{m}, \quad k=1,\cdots,m
\]
and  
\[
\cos\left(\frac{k\pi}{m}\right) < \cos\left(\frac{k\pi}{m+1}\right)< \cos\left(\frac{(k-1)\pi}{m}\right), k=1,\cdots,m.
\]
Recalling that the roots of $U_{m}(y)$ are $\cos \frac{k\pi}{m+1}, \ k=1,2,\cdots, m$, we are now ready to estimate $f(y)$. We have 
\[
f\left(\cos\left(\frac{\pi}{m+1}\right)\right)> 0, \ f\left(\cos\left(\frac{2\pi}{m+1}\right)\right)< 0, \ f\left(\cos\left(\frac{3\pi}{m+1}\right)\right)>0, \ \cdots
\]
Thus, the solutions $\{y_k\}$ to $f(y)=0$ satisfy that, except for only at most two $y_k$'s, after reindexation,  
\[
y_k=\cos\theta_k, \quad \theta_k \in \left( \frac{k\pi}{m+1}, \frac{(k+1)\pi}{m+1} \right),\ k=1, \cdots, m-1.   
\]
Then, similarly to the discussions in Step 1 of the proof of Theorem \ref{thm:eigenvaluethm1}, we can reindex the $\lambda_r$'s to obtain
\[
 \lambda_{1}^l < \lambda_{2}^l<\cdots< \lambda_{m-1}^l\leq \min\{\alpha_1, \alpha_2\}\leq  \max\{\alpha_1, \alpha_2\}\leq \lambda_{m-1}^r< \cdots< \lambda_{2}^{r}< \lambda_{1}^{r}
\]
with 
\begin{equation*}
\cos\left(\frac{(k+1)\pi}{m+1}\right)\leq \min\left\{y\left(\lambda_k^l\right), y\left(\lambda_k^r\right)\right\}\leq \max\left\{y\left(\lambda_k^l\right), y\left(\lambda_k^r\right)\right\}\leq \cos\left(\frac{k\pi}{m+1}\right)
\end{equation*}
for $k=1,2,\cdots, m$. Analogously, we can prove the result for the case when $m$ is an even number. 

\medskip
\noindent \textbf{Step 2.} Similarly to Step 2 in the proof of Theorem \ref{thm:eigenvaluethm1}, by considering principal submatrices of $A_{2m}^{(a,b)}$ and utilizing the result in Step 1 together with the Cauchy interlacing theorem, we can prove the statement.  
\end{proof}

\section{Eigenvectors of tridiagonal $2$-Toeplitz matrices with perturbations on the diagonal corners} \label{sect3}
This section serves to construct the formula of the eigenvectors of tridiagonal $2$-Toeplitz matrices with perturbations on the diagonal corners through a general strategy. It generalizes the results obtained in \cite{gover1994eigenproblem}.

\subsection{Preliminaries}
We start by introducing the following two families of polynomials. 
\begin{defi}\label{defi:generalqmpm1}
	We define the two families of polynomials $q_{k}^{(\xi_p, \xi_{q})}, p_{k}^{(\xi_p, \xi_{q})}$ by 
	\[
	q_{0}^{(\xi_p, \xi_{q})}(\nu)=\xi_{q}, \qquad p_{0}^{\xi_p, \xi_{p}}(\nu)=\xi_{p},
	\]
	and the recurrence formulas
	\begin{equation}\label{equ:recurrencerelation5}
		q_{k}^{(\xi_p, \xi_{q})}(\nu)=\nu p_{k-1}^{(\xi_p, \xi_{q})}(\nu)-q_{k-1}^{(\xi_p, \xi_{q})}(\nu)
	\end{equation}
	and
	\begin{equation}\label{equ:recurrencerelation6}
		p_{k}^{(\xi_p, \xi_{q})}(\nu) = q_{k}^{(\xi_p, \xi_{q})}(\nu) - \zeta p_{k-1}^{(\xi_p, \xi_{q})}(\nu),
	\end{equation}
	where 
	\begin{equation}\label{equ:zeta2}
		\zeta = \frac{\gamma_{2} \beta_{2}}{\gamma_{1} \beta_{1}}.
	\end{equation}
\end{defi}

\medskip
Then we observe that the recurrence formulas (\ref{equ:recurrencerelation5}) and (\ref{equ:recurrencerelation6}) can be simplified. 
\begin{prop}\label{thm:recurrencerelation1}
	If $p_k^{(\xi_p, \xi_{q})}(\nu)$ and $q_k^{(\xi_p, \xi_{q})}(\nu)$ satisfy (\ref{equ:recurrencerelation5}) and (\ref{equ:recurrencerelation6}) respectively, then
	\begin{equation}\label{equ:recurrencerelation3}
		p_{k+1}^{(\xi_p, \xi_{q})}(\nu)=[\nu-(1+\zeta)] p_k^{(\xi_p, \xi_{q})}(\nu)-\zeta p_{k-1}^{(\xi_p, \xi_{q})}(\nu)
	\end{equation}
	and
	\begin{equation}\label{equ:recurrencerelation4}
		q_{k+1}^{(\xi_p, \xi_{q})}(\nu)=[\nu-(1+\zeta)] q_k^{(\xi_p, \xi_{q})}(\nu)-\zeta q_{k-1}^{(\xi_p, \xi_{q})}(\nu)
	\end{equation}
	with
	\begin{equation}\label{equ:recurrenceinitial1}
 \begin{aligned}
		&p_0^{(\xi_p, \xi_{q})}(\nu)=\xi_p, \quad p_1^{(\xi_p, \xi_{q})}(\nu)=\left(\nu-\zeta\right)\xi_{p}-\xi_{q}, \\
  &q_0^{(\xi_p, \xi_{q})}(\nu)=\xi_{q}, \quad q_1^{(\xi_p, \xi_{q})}(\nu)=\nu\xi_{p}-\xi_{q},
  \end{aligned}
	\end{equation}
	where $\zeta$ is defined in (\ref{equ:zeta2}).
\end{prop}
\begin{proof}
 From (\ref{equ:recurrencerelation5}) and (\ref{equ:recurrencerelation6}), we have 
 \[
 	p_{k+1}^{(\xi_p, \xi_{q})}(\nu)=(\nu-\zeta)p_{k}^{(\xi_p, \xi_{q})}(\nu)-q_{k}^{(\xi_p, \xi_{q})}(\nu),
 \]
which gives
 \[
q_{k}^{(\xi_p, \xi_{q})}(\nu) =  (\nu-\zeta)p_{k}^{(\xi_p, \xi_{q})}(\nu) - p_{k+1}^{(\xi_p, \xi_{q})}(\nu).
 \]
 Substituting the above identity into (\ref{equ:recurrencerelation6}) yields (\ref{equ:recurrencerelation3}). Similarly, from (\ref{equ:recurrencerelation5}), it follows that
 \[
 	\nu p_{k-1}^{(\xi_p, \xi_{q})}(\nu)= q_{k}^{(\xi_p, \xi_{q})}(\nu)+ q_{k-1}^{(\xi_p, \xi_{q})}(\nu),
 \]
which together with (\ref{equ:recurrencerelation6}) yields (\ref{equ:recurrencerelation4}). By Definition \ref{defi:generalqmpm1}, (\ref{equ:recurrenceinitial1}) can be easily verified.
\end{proof}

\subsection{Normalisation}\label{section:normalization1}
To help to explain the skin effect later, in this section we 
normalise the polynomials $p_{k}^{(\xi_p, \xi_{q})}(\nu), q_{k}^{(\xi_p, \xi_{q})}(\nu)$. This can be achieved by setting
\begin{equation}\label{equ:defineofmu1}
\mu=\frac{\nu-\left(1+\beta^2\right)}{2\beta}
\end{equation}
with
$$
\beta^2=\frac{\beta_2 \gamma_2}{\beta_1 \gamma_1},
$$
and
\begin{equation}\label{equ:normalizepolynomial1}
\widehat p_k^{(\xi_{p}, \xi_{q})}(\mu)=\frac{1}{\beta^k} p_k^{(\xi_{p}, \xi_{q})}\left(1+2\beta \mu+\beta^2\right), \ \widehat q_k^{(\xi_{p}, \xi_{q})}(\mu)=\frac{1}{\beta^k} q_k^{(\xi_{p}, \xi_{q})}\left(1+2\beta \mu+\beta^2\right). 
\end{equation}
Thus from (\ref{equ:recurrencerelation3}) and (\ref{equ:recurrencerelation4}), we get respectively,
\begin{equation}\label{equ:Chebyshevrecurrence1}
	\widehat p_{k+1}^{(\xi_{p}, \xi_{q})}(\mu)=2\mu \widehat p_k^{(\xi_{p}, \xi_{q})}(\mu)-\widehat p_{k-1}^{(\xi_{p}, \xi_{q})}(\mu),
\end{equation}
and
\begin{equation}\label{equ:Chebyshevrecurrence2}
	\widehat q_{k+1}^{(\xi_{p}, \xi_{q})}(\mu)=2\mu \widehat q_k^{(\xi_{p}, \xi_{q})}(\mu)-\widehat q_{k-1}^{(\xi_{p}, \xi_{q})}(\mu).
\end{equation}
Equations (\ref{equ:Chebyshevrecurrence1}) and (\ref{equ:Chebyshevrecurrence2}) are the Chebyshev three point recurrence formula. Also, from (\ref{equ:recurrenceinitial1}) the initial polynomials are given by
\begin{equation}\label{equ:recurrenceinitial2}
\begin{aligned}
&\widehat p_{0}^{(\xi_{p}, \xi_{q})}(\mu) =\xi_{p},\quad  \widehat p_{1}^{(\xi_{p}, \xi_{q})}(\mu) =2\mu\xi_{p}+ \frac{\xi_{p}-\xi_{q}}{\beta}, \\
 & \widehat q_{0}^{(\xi_{p}, \xi_{q})}(\mu) =\xi_{q},\quad  \widehat q_{1}^{(\xi_{p}, \xi_{q})}(\mu) =(2\mu+\beta)\xi_{p}+\frac{\xi_{p}-\xi_{q}}{\beta}. 
\end{aligned}
\end{equation}

\subsection{Eigenvectors of $A_{2m+1}^{(a,b)}$ and $A_{2m}^{(a,b)}$}
In this section, we present a formula for the eigenvectors of tridiagonal $2$-Toeplitz matrices with perturbations on diagonal corners through a direct construction. This formula shows that, even though some of the elements of the $2$-Toeplitz matrix were perturbed, the structure of the eigenvectors is similar to the one in \cite{gover1994eigenproblem}. 

\bigskip
\begin{thm}\label{thm:eigenvectoroddmatrixtwoperturb2}
	The eigenvector of $A_{2m+1}^{(a, b)}$ in (\ref{equ:oddmatrixtwoperturb1}) associated with the eigenvalue $\lambda_r$ is given by
	\begin{equation}
		\begin{aligned}\label{equ:eigenvectoroddmatrixtwoperturb3}
			\mathbf{x}= & \left(\widehat q_{0}^{(\xi_p, \xi_{q})}\left(\mu_r\right),-\frac{1}{\beta_1}\left(\alpha_1-\lambda_r\right) \widehat p_0^{(\xi_p, \xi_{q})}\left(\mu_r\right), s \widehat q_1^{(\xi_p, \xi_{q})}\left(\mu_r\right),  -\frac{1}{\beta_1} s\left(\alpha_1-\lambda_r\right) \widehat p_1^{(\xi_p, \xi_{q})}\left(\mu_r\right),\right. \\
			&\qquad  \left. \ldots,  -\frac{1}{\beta_1} s^{m-1}\left(\alpha_1-\lambda_r\right) \widehat p_{m-1}^{(\xi_p, \xi_{q})}\left(\mu_r\right), s^m \widehat q_m^{(\xi_p, \xi_{q})}\left(\mu_r\right)\right)^{\top},
		\end{aligned}
	\end{equation}
 where $\top$ denotes the transpose and  $\widehat q_{k}^{(\xi_p, \xi_{q})}\left(\cdot\right), \widehat p_{k}^{(\xi_p, \xi_{q})}\left(\cdot\right)$ are defined as in (\ref{equ:normalizepolynomial1}) with $\xi_{q}=(\alpha_1-\lambda_r), \xi_{p}= (\alpha_1+a-\lambda_{r})$, and $s, \mu_r$ are  respectively given by
	\begin{equation}\label{equ:defiofstwoperturb2}
		s= \sqrt{\frac{\gamma_{1}\gamma_{2}}{\beta_{1}\beta_2}}, \quad \mu_r = \frac{\left(\alpha_1-\lambda_r\right)\left(\alpha_2-\lambda_r\right)-(\gamma_1\beta_1+\gamma_2\beta_2)}{2\sqrt{\gamma_1\beta_1\gamma_2\beta_2}}.
	\end{equation}
\end{thm}

\begin{proof}
We first demonstrate that the eigenvector of $A_{2m+1}^{(a,b)}$ in (\ref{equ:oddmatrixtwoperturb1}) associated with the eigenvalue $\lambda_r$ has the following form: \begin{align}\label{equ:eigenvectoroddmatrixtwoperturb1}
			\mathbf{x}= & \left(q_{0}^{(\xi_p, \xi_{q})}\left(\nu_r\right),-\frac{1}{\beta_1}\left(\alpha_1-\lambda_r\right) p_0^{(\xi_p, \xi_{q})}\left(\nu_r\right), \left(\frac{\gamma_1}{\beta_2}\right) q_1^{(\xi_p, \xi_{q})}\left(\nu_r\right), \right. \nonumber \\
   & -\frac{1}{\beta_1} \left(\frac{\gamma_1}{\beta_2}\right)\left(\alpha_1-\lambda_r\right) p_1^{(\xi_p, \xi_{q})}\left(\nu_r\right), \ldots, -\frac{1}{\beta_1} \left(\frac{\gamma_1}{\beta_2}\right)^{m-1}\left(\alpha_1-\lambda_r\right) p_{m-1}^{(\xi_p, \xi_{q})}\left(\nu_r\right),\nonumber  \\ 
    &\qquad \left.
   \left(\frac{\gamma_1}{\beta_2}\right)^m q_m^{(\xi_p, \xi_{q})}\left(\nu_r\right)\right)^{\top},
		\end{align}
	where $q_{k}^{(\xi_p, \xi_{q})}\left(\cdot \right), p_{k}^{(\xi_p, \xi_{q})}\left(\cdot \right)$ are the polynomials defined as in Definition \ref{defi:generalqmpm1} with $\xi_{q}=(\alpha_1-\lambda_r), \xi_{p}= (\alpha_1+a-\lambda_{r})$, and $\nu_r$ is given by 
\begin{equation}\label{equ:eigenvectoroddmatrixtwoperturb2}
    \nu_r=\frac{\left(\alpha_1-\lambda_r\right)\left(\alpha_2-\lambda_r\right)}{\gamma_1 \beta_1}.
\end{equation}
To prove (\ref{equ:eigenvectoroddmatrixtwoperturb1}), we consider 
\[
\left(A_{2m+1}-\lambda_{r} I\right) \vect x=0,
\]
where 
\[
\vect x=\left(x_1, -\frac{1}{\beta_1}(\alpha_1-\lambda_{r})x_2, \left(\frac{\gamma_1}{\beta_2}\right)x_3, \cdots, -\frac{1}{\beta_1} \left(\frac{\gamma_1}{\beta_2}\right)^{m-1}\left(\alpha_1-\lambda_r\right)x_{2m},  \left(\frac{\gamma_1}{\beta_2}\right)^m x_{2m+1}\right)^{\top}.
\] 
Considering the first row, we can choose
\[
	x_1 =(\alpha_1-\lambda_r), \quad x_2 =  (\alpha_{1}+a -\lambda_{r}).
\]
Then by the second row, we have 
	\[
	\gamma_{1} x_1 -\frac{1}{\beta_{1}} (\alpha_{1}-\lambda_{r})(\alpha_{2}-\lambda_{r})x_2 +\gamma_1x_3 =0,   
	\]
 which gives 
	\[
	x_3 =  \nu_{r}x_2-x_1.
	\]
	The third row is 
	\[
	-\frac{\gamma_{2}}{\beta_{1}} (\alpha_{1}-\lambda_{r})x_2 +(\alpha_{1}-\lambda_{r})\left(\frac{\gamma_1}{\beta_2}\right)x_3 -(\alpha_{1}-\lambda_{r})\left(\frac{\gamma_1}{\beta_2}\right)x_4=0,
	\]
	and thus, 
	\begin{align*}
	x_4 = - \frac{\beta_{2}\gamma_{2}}{\beta_{1}\gamma_{1}}x_2+x_3= x_3 -\zeta x_2,
	\end{align*}
where $\zeta = \frac{\beta_{2}\gamma_{2}}{\beta_{1}\gamma_{1}}$. Continuing the process, we can easily verify that 
\begin{align*}
x_{2k+1} &= \nu_{r} x_{2k}- x_{2k-1}, \quad k=1, \cdots, m,\\
x_{2k}&=x_{2k-1} -\zeta x_{2k-2},\quad k=2,\cdots, m.
\end{align*}
By the definition of $q_{k}^{(\xi_p, \xi_{q})}(\nu_{r}), p_{k}^{(\xi_p, \xi_{q})}(\nu_{r})$ in Definition \ref{defi:generalqmpm1}, we note that 
\begin{align*}
x_{2k+1} &= q_{k}^{(\xi_p, \xi_{q})}(\nu_{r}), \quad  k=0,\cdots, m,\\
x_{2k} &= p_{k}^{(\xi_p, \xi_{q})}(\nu_{r}), \quad  k=1,\cdots, m,
\end{align*}
with $\xi_{q}=(\alpha_1-\lambda_r), \xi_{p}= (\alpha_1+a-\lambda_{r})$. This proves (\ref{equ:eigenvectoroddmatrixtwoperturb1}). 

Finally, by (\ref{equ:defineofmu1}) and (\ref{equ:normalizepolynomial1}), we can write (\ref{equ:eigenvectoroddmatrixtwoperturb1}) as (\ref{equ:eigenvectoroddmatrixtwoperturb2}). This completes the proof. 
\end{proof}

In the same manner, we have the following theorem for the eigenvectors of $A_{2m}^{(a, b)}$. 

\begin{thm}\label{thm:eigenvectorevenmatrixtwoperturb2}
	The eigenvector of $A_{2m}^{(a,b)}$ in (\ref{equ:evenmatrixtwoperturb1}) associated with the eigenvalue $\lambda_r$ is given by
	\begin{equation}
		\begin{aligned}\label{equ:eigenvectorevenmatrixtwoperturb3}
			\mathbf{x}= & \left(\widehat q_{0}^{(\xi_p, \xi_{q})}\left(\mu_r\right),-\frac{1}{\beta_1}\left(\alpha_1-\lambda_r\right) \widehat p_0^{(\xi_p, \xi_{q})}\left(\mu_r\right), s \widehat q_1^{(\xi_p, \xi_{q})}\left(\mu_r\right),  -\frac{1}{\beta_1} s\left(\alpha_1-\lambda_r\right) \widehat p_1^{(\xi_p, \xi_{q})}\left(\mu_r\right), \right. \\
			&\qquad  \left.  \ldots,-\frac{1}{\beta_1} s^{m-1}\left(\alpha_1-\lambda_r\right) \widehat p_{m-1}^{(\xi_p, \xi_{q})}\left(\mu_r\right)\right)^{\top},
		\end{aligned}
	\end{equation}
	where $\xi_{q}=(\alpha_1-\lambda_r), \xi_{p}= (\alpha_1+a-\lambda_{r})$, $s$ and $\mu_r$ are defined as in (\ref{equ:defiofstwoperturb2}).
\end{thm}

\begin{figure}
    \centering
    \includegraphics[width=0.5\textwidth]{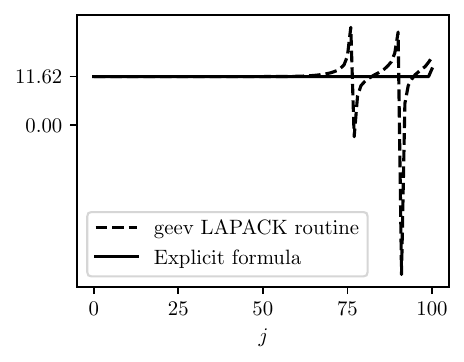}
    \caption{Theorems \ref{thm:eigenvectoroddmatrixtwoperturb2} and \ref{thm:eigenvectorevenmatrixtwoperturb2} outperform numerical solvers. For $\alpha_1=1,\alpha_2=2, \beta_1=3,\beta_2=4,\gamma_1=4,\gamma_2=5, a=9, b=10$ and $m=50$, the numerical solvers produce worse outputs due to floating point arithmetic limitations. For the eigenvalue $\lambda_{2m-1}\approx11.6217$, the figure shows $A_{2m+1}v_{2m-1}^{(j)}/v_{2m-1}^{(j)}$ (where $v^{(j)}$ is the $j$-th entry of the vector) for $v_{2m-1}$ computed with standard numerical routines (dashed line) and computed with Theorem \ref{thm:eigenvectoroddmatrixtwoperturb2} (solid line). The expected result is a constant line at $11.621$.}
    \label{fig: eigenvectors numeric vs exact}
\end{figure}

\begin{remark}
	When $a=0$, choosing $\xi_{p}=\xi_{q}=1$ in the above two theorems, the results are reduced to the ones obtained in \cite{gover1994eigenproblem}. Thus, the method and findings presented here are generalisations of \cite{gover1994eigenproblem}. We note that the authors of \cite{barnett2023effects} also proposed a way  to derive the eigenvectors of some tridiagonal $2$-Toeplitz matrices with perturbations on four corners, but their method cannot be easily adapted to prove the condensation of eigenvectors of 
 the perturbed tridiagonal $K$-Toeplitz matrices arising in the non-Hermitian skin effect.
\end{remark}

\begin{remark}
Note that the crucial idea in the proofs of Theorems \ref{thm:eigenvectoroddmatrixtwoperturb2} and \ref{thm:eigenvectorevenmatrixtwoperturb2} is that when the $2$-Toeplitz structure appears in certain parts of the matrix, then we can start to construct these parts of the eigenvector by the polynomials from the recurrence formula in Definition \ref{defi:generalqmpm1} and ignore the irregular parts. This strategy can be applied to construct the eigenvectors of $K$-Toeplitz matrix with more complicated perturbations.
\end{remark}

\begin{remark}
By (\ref{equ:recurrenceinitial2}), in Theorems \ref{thm:eigenvectoroddmatrixtwoperturb2} and \ref{thm:eigenvectorevenmatrixtwoperturb2}, the initial values of $\widehat p_{k}^{(\xi_{p}, \xi_{q})}(\mu_r)$ and $\widehat q_{k}^{(\xi_{p}, \xi_{q})}(\mu_r)$ are
\begin{equation}\label{equ:recurrenceinitial2b}
\begin{array}{ll}
	\widehat p_{0}^{(\xi_{p}, \xi_{q})}(\mu_r) =(\alpha_1-\lambda_r),& \widehat p_{1}^{(\xi_{p}, \xi_{q})}(\mu_r) =2\mu_r(\alpha_1-\lambda_r) -\frac{a}{\beta}, \\
	 \widehat q_{0}^{(\xi_{p}, \xi_{q})}(\mu_r) =(\alpha_1+a-\lambda_{r}),&\widehat q_{1}^{(\xi_{p}, \xi_{q})}(\mu_r) =(2\mu_r+\beta)(\alpha_1-\lambda_r) -\frac{a}{\beta}.
\end{array}
\end{equation}
\end{remark}

\begin{remark}
    Figure \ref{fig: eigenvectors numeric vs exact} shows that the formulas obtained in Theorems \ref{thm:eigenvectoroddmatrixtwoperturb2} and \ref{thm:eigenvectorevenmatrixtwoperturb2} outperform the standard numerical routines. This is already noticeable in relatively small matrices of size $101\times 101$ and is due to floating point arithmetic limitations. 
\end{remark}

\subsection{Other representations}
To help to demonstrate the existence of interface modes later, we parameterise $A_{2m+1}^{(a,b)}$ in (\ref{equ:oddmatrixtwoperturb1}) as $A_{2m+1}^{(a,b)}(\alpha_{1}, \beta_{1}, \gamma_{1}, \alpha_{2}, \beta_{2}, \gamma_{2})$.  Throughout the rest of the paper, if not specified, $A_{2m+1}^{(a,b)}$ is $A_{2m+1}^{(a,b)}(\alpha_{1}, \beta_{1}, \gamma_{1}, \alpha_{2}, \beta_{2}, \gamma_{2})$. 

Note that
\begin{align*}
A_{2m+1}^{(b,a)}(\alpha_{1},\gamma_{2}, \beta_{2}, \alpha_{2}, \gamma_{1}, \beta_{1}) = R_{2m+1} A_{2m+1}^{(a,b)}(\alpha_{1}, \beta_{1}, \gamma_{1}, \alpha_{2}, \beta_{2}, \gamma_{2})R_{2m+1},
\end{align*}
where 
\begin{equation}\label{equ:antidiagonalidentity1}
R_{2m+1} = \begin{pmatrix}
	0&0&0&\cdots&0&1\\
	0&0&0&\cdots&1&0\\
	\vdots&\vdots&\vdots&\ddots&\vdots&\vdots\\
	0&0&1&\cdots&0&0\\
	0&1&0&\cdots&0&0\\
	1&0&0&\cdots&0&0	
\end{pmatrix}.
\end{equation}
It follows that the eigenvalues $\{\lambda_{r}\}$ of $A_{2m+1}^{(b,a)}(\alpha_{1},\gamma_{2}, \beta_{2}, \alpha_{2}, \gamma_{1}, \beta_{1}) $ are the same as those of $A_{2m+1}^{(a,b)}(\alpha_{1}, \beta_{1}, \gamma_{1}, \alpha_{2}, \beta_{2}, \gamma_{2})$ and the eigenvectors of $A_{2m+1}^{(b,a)}(\alpha_{1},\gamma_{2}, \beta_{2}, \alpha_{2}, \gamma_{1}, \beta_{1})$ are of the form  
	\begin{align}
		\mathbf{x}= & \left( s^m \widehat q_m^{(\xi_p, \xi_{q})}\left(\mu_r\right), -\frac{1}{\beta_1} s^{m-1}\left(\alpha_1-\lambda_r\right) \widehat p_{m-1}^{(\xi_p, \xi_{q})}\left(\mu_r\right),\cdots,-\frac{1}{\beta_1} s\left(\alpha_1-\lambda_r\right) \widehat p_1^{(\xi_p, \xi_{q})}\left(\mu_r\right),  \right. \nonumber \\
		 &\left.  \qquad s \widehat q_1^{(\xi_p, \xi_{q})}\left(\mu_r\right), -\frac{1}{\beta_1}\left(\alpha_1-\lambda_r\right) \widehat p_0^{(\xi_p, \xi_{q})}\left(\mu_r\right), \widehat q_{0}^{(\xi_p, \xi_{q})}\left(\mu_r\right)\right)^{\top} , \label{equ:eigenvectorreverseoddmatrixtwoperturb3}
	\end{align}
where $\xi_{q}=(\alpha_1-\lambda_r), \xi_{p}= (\alpha_1+a-\lambda_{r})$ and $s$, $\mu_r$ are defined as in (\ref{equ:defiofstwoperturb2}).

\section{Exponential decay and localisation of eigenvectors}
\label{sect4}
\subsection{Exponential decay of the eigenvectors}
In this section, we demonstrate the exponential decay for the entries of the eigenvectors of matrices $A_{2m+1}^{(a,b)}, A_{2m}^{(a,b)}$. To do so, by Theorems \ref{thm:eigenvectoroddmatrixtwoperturb2} and \ref{thm:eigenvectorevenmatrixtwoperturb2}, the only thing left is to control the polynomials $ \widehat p_{k}^{(\xi_p, \xi_{q})}\left(\mu_r\right)$ and $ \widehat q_{k}^{(\xi_p, \xi_{q})}\left(\mu_r\right)$. This requires information on the eigenvalues and the Chebyshev polynomials in Section \ref{section:eigenvalueanalysis1}. The main theorem is presented below.

\begin{thm}\label{thm:skineffect1}
Except for at most $11$ eigenvalues $\{\lambda_r\}$ of $A_{2 m+1}^{(a, b)}$, the corresponding eigenvectors $\vect x$ in Theorem \ref{thm:eigenvectoroddmatrixtwoperturb2} satisfy that 
\begin{equation}\label{equ:skineffectequ1}
\babs{\vect x^{(j)}}\leq Mj \left(\sqrt{\frac{\gamma_{1}\gamma_{2}}{\beta_{1}\beta_{2}}}\right)^{\lfloor\frac{j-1}{2}\rfloor}
\end{equation}
for some constant $M>0$ independent of the $\lambda_r$'s, where $\vect x^{(j)}$ is the $j$-th component of ${\vect x}_r$. The estimate \eqref{equ:skineffectequ1} holds also for eigenvectors $\vect x$ of $A_{2 m}^{(a,b)}$ associated with the eigenvalue $\lambda_r$ in Theorem \ref{thm:eigenvectorevenmatrixtwoperturb2}, except for at most $12$ $r$'s.  
\end{thm}
\begin{proof}
By Theorems \ref{thm:eigenvaluethm1} and \ref{thm:eigenvaluethm2}, all the $\lambda_{r}$'s of $A_{2m+1}^{(a,b)}$ and $A_{2m}^{(a, b)}$ are real numbers, and except for at most $11$ $\lambda_r$'s of $A_{2m+1}^{(a,b)}$, we have 
\begin{align}\label{equ:proofskineffect1}
\mu_r=\frac{(\lambda_{r}-\alpha_{1})(\lambda_r-\alpha_{2})-\gamma_{1}\beta_{1}-\gamma_2\beta_{2} }{2\sqrt{\gamma_{1}\beta_{1}\gamma_{2}\beta_{2}}}=\cos \theta_{r}
\end{align}
for certain $\theta_r\in [0, \pi]$. Now we are going to demonstrate that $ \widehat p_{k}^{(\xi_p, \xi_{q})}\left(\mu_r\right)$ and $ \widehat q_{k}^{(\xi_p, \xi_{q})}\left(\mu_r\right)$ are bounded for $\mu_r=\cos\theta_r$. The idea is to represent them with Chebyshev polynomials, although they are not Chebyshev polynomials straightforwardly. To do so, we separate $ \widehat p_{k}^{(\xi_p, \xi_{q})}\left(x\right)$ as follows
\[
 \widehat p_{k}^{(\xi_p, \xi_{q})}\left(x\right) = u_k(x)+v_{k-1}(x)
\]
with 
\[
\begin{array}{ll}
u_0(x) = (\alpha_1-\lambda_r), &\ u_1(x)=2x(\alpha_1-\lambda_r), \\
v_{-1}(x)=0,&\ v_{0}(x) = -\frac{a}{\beta}.
\end{array}
\]
Note that by (\ref{equ:recurrenceinitial2b}), we have $\widehat p_{0}^{(\xi_p, \xi_{q})}\left(\mu_r\right) = u_0(\mu_r)+v_{-1}(\mu_r), \widehat p_{1}^{(\xi_p, \xi_{q})}\left(\mu_r\right) = u_1(\mu_r)+v_{0}(\mu_r)$. By (\ref{equ:Chebyshevrecurrence1}), it is not hard to check that
\begin{align*}
&u_{k+1}(x) = 2x u_{k}(x) - u_{k-1}(x),\\
&v_{k+1}(x) = 2x v_{k}(x) - v_{k-1}(x),
\end{align*}
and 
\[
\begin{array}{ll}
u_0(x) = (\alpha_1-\lambda_r), &\ u_1(x)=2x(\alpha_1-\lambda_r), \\
v_{0}(x) = -\frac{a}{\beta}, &\ v_1(x)= 2x \left( -\frac{a}{\beta}\right).
\end{array}
\]
Thus both the $u_k(x), v_k(x)$ are Chebyshev polynomials of the second kind after scaling. By (\ref{equ:boundofchebyshev1}), we have for $-1 \leq x \leq 1$ and $k=0,1,\cdots, $
\[
 \babs{\widehat p_{k}^{(\xi_p, \xi_{q})}\left(x\right)}\leq \babs{ u_k(x)}+\babs{v_{k-1}(x)}\leq \left(k+1\right)\babs{\alpha_1-\lambda_r} + k \babs{\frac{a}{\beta}}.
\]
It is not hard to see that, for $\lambda_r$ satisfying (\ref{equ:proofskineffect1}),  $|\alpha_1-\lambda_r|$ and $\babs{\frac{a}{\beta}}$ are uniformly bounded. That is, 
\[
\babs{\widehat p_{k}^{(\xi_p, \xi_{q})}\left(x\right)}\leq k\tilde{M}, \quad x\in [-1,1],
\]
for some $\tilde{M}>0$. 

To demonstrate the boundness of the quantities $ \widehat q_{k}^{(\xi_p, \xi_{q})}\left(\mu_r\right)$ we separate them as 
\[
\widehat q_{k}^{(\xi_p, \xi_{q})}\left(x\right) = u_k(x)+w_k(x)+ v_{k-1}(x)
\]
with 
\begin{align*}
	\begin{array}{ll}
u_0(x) = (\alpha_1-a-\lambda_r), &u_1(x)=2x(\alpha_1-a-\lambda_r), \\
w_0(x) = 2a, & w_1(x)=x(2a), \\
 v_{-1}(x)=0,& v_{0}(x) =(\alpha_1 -\lambda_r)\beta+\frac{-a}{\beta}.
 \end{array}
\end{align*}
It is not hard to see that $u_k(x), v_k(x)$ are Chebyshev polynomials of the second kind after scaling and $w_k(x)$ are Chebyshev polynomials of the first kind after scaling. In the same fashion, by (\ref{equ:boundofchebyshev1}) we can show that $\babs{\widehat q_{k}^{(\xi_p, \xi_{q})}\left(\mu_r\right)}$ are bounded by $k\tilde{M}$ for some $\tilde{M}$>0. Considering the formula of the eigenvector $\vect x$ in Theorem \ref{thm:eigenvectoroddmatrixtwoperturb2}, it is enough to demonstrate (\ref{equ:skineffectequ1}). The result for $A_{2m}^{(a,b)}$ follows from Theorem \ref{thm:eigenvaluethm2}. This completes the proof.
\end{proof}

\begin{remark}
It should be noted that the method and results presented in this section, as well as the preceding ones, can be extended to tridiagonal $K$-Toeplitz matrices with certain perturbations. This generalisation will be presented in future work.
\end{remark}

\subsection{Localised eigenvectors} \label{subsect4}
In this section, we present a theorem for demonstrating the localisation of the coefficients of the eigenvectors of block matrices with mirrored tridiagonal $2$-Toeplitz blocks. We define the matrix $C_{2m+1, 2m+1}$ of order $4m+2$ by
\begin{equation}\label{equ:twotoeplitzmatrix1}
\begin{aligned}
	C_{2m+1, 2m+1}^{(a,b)}
	=\left(
	\begin{array}{ll}
			G_{11} & G_{12} \\
			 G_{21}& G_{22}  
	\end{array}
	\right),
\end{aligned}
\end{equation}
where $G_{11}=R_{2m+1} A_{2m+1}^{(0,a)} R_{2m+1}$ with $R_{k}$ being defined as in (\ref{equ:antidiagonalidentity1}), $G_{22} = A_{2m+1}^{(0,b)}$, and  
\[
 G_{12}  = \begin{pmatrix}
	0&0&\cdots &0\\
	\vdots&\vdots&\vdots& \vdots\\
	0&0&\cdots &0\\
	\gamma_{2}&0&\cdots&0
\end{pmatrix}, \quad  G_{21}  = \begin{pmatrix}
0&\cdots&0 &\gamma_{2}\\
\vdots&\vdots&\vdots& \vdots\\
0&\cdots &0&0\\
0&\cdots&0&0
\end{pmatrix}.
\] 

Below is the main theorem for the interface modes. For the matrix $C_{2m, 2m}$, one can demonstrate similar results in the same manner. 

\begin{thm}\label{thm:eigenvectorlocalizedmode1}
	Let $\{\lambda_{r}\}$ be the eigenvalues of $C_{2m+1, 2m+1}$ in (\ref{equ:twotoeplitzmatrix1})  and define
	\begin{align*}
	&s = \sqrt{\frac{\gamma_{1}\gamma_{2}}{\beta_{1}\beta_{2}}}, \quad  \mu_{r}= \frac{(\lambda_r-\alpha_{1})(\lambda_{r}-\alpha_{2})-(\gamma_{1}\beta_1+\gamma_{2}\beta_2)}{2 \sqrt{\gamma_{1}\beta_1 \gamma_{2}\beta_2}}.	
	\end{align*}
The corresponding eigenvector of $C_{2m+1, 2m+1}$ is given by
	\begin{equation}
		\vect x = \begin{pmatrix}
			\vect x_1\\
			 \vect x_2\\
			  \vect x_3\end{pmatrix}.
	\end{equation}
The $\vect x_1$ part has the form
\begin{align*}
	\vect x_1 =& \left( \left(s\right)^m \widehat q_m^{(\xi_p, \xi_{q})}\left(\mu_r\right), -\frac{1}{\beta_1} \left(s\right)^{m-1}\left(\alpha_1-\lambda_r\right) \widehat p_{m-1}^{(\xi_p, \xi_{q})}\left(\mu_r\right),\cdots,-\frac{1}{\beta_1} s\left(\alpha_1-\lambda_r\right) \widehat p_1^{(\xi_p, \xi_{q})}\left(\mu_r\right),  \right. \nonumber \\
	&\left.  \qquad s \widehat q_1^{(\xi_p, \xi_{q})}\left(\mu_r\right), -\frac{1}{\beta_1}\left(\alpha_1-\lambda_r\right) \widehat p_0^{(\xi_p, \xi_{q})}\left(\mu_r\right), \widehat q_{0}^{(\xi_p, \xi_{q})}\left(\mu_r\right)\right)^{\top},
\end{align*}
or
\begin{align*}
\vect x_1=& \left( -\frac{1}{\beta_1} \left(s\right)^{m-1}\left(\alpha_2-\lambda_r\right) \widehat p_{m-1}^{(\xi_p, \xi_{q})}\left(\mu_r\right),\cdots,-\frac{1}{\beta_1} s\left(\alpha_2-\lambda_r\right) \widehat p_1^{(\xi_p, \xi_{q})}\left(\mu_r\right),  \right. \nonumber \\
	&\left.  \qquad s \widehat q_1^{(\xi_p, \xi_{q})}\left(\mu_r\right), -\frac{1}{\beta_1}\left(\alpha_2-\lambda_r\right) \widehat p_0^{(\xi_p, \xi_{q})}\left(\mu_r\right), \widehat q_{0}^{(\xi_p, \xi_{q})}\left(\mu_r\right)\right)^{\top},
\end{align*}
where $\xi_{q}=C, \xi_{p}=C$ for a certain constant $C$.  

\medskip
\noindent The $\vect x_3$ part has the form 
\begin{equation*}
\begin{aligned}
	 	\vect x_3 = &\left( \widehat q_{0}^{(\xi_p, \xi_{q})}\left(\mu_r\right),-\frac{1}{\beta_1}\left(\alpha_1-\lambda_r\right) \widehat p_0^{(\xi_p, \xi_{q})}\left(\mu_r\right), s \widehat q_1^{(\xi_p, \xi_{q})}\left(\mu_r\right),  -\frac{1}{\beta_1} s\left(\alpha_1-\lambda_r\right) \widehat p_1^{(\xi_p, \xi_{q})}\left(\mu_r\right), \right. \\
		&\qquad  \left.  \ldots, -\frac{1}{\beta_1} (s)^{m-1}\left(\alpha_1-\lambda_r\right) \widehat p_{m-1}^{(\xi_p, \xi_{q})}\left(\mu_r\right), (s)^m \widehat q_m^{(\xi_p, \xi_{q})}\left(\mu_r\right)\right )^{\top}
	\end{aligned}
\end{equation*}
or   
\[
\begin{aligned}
	\vect x_3 =& \left( \widehat q_{0}^{(\xi_p, \xi_{q})}\left(\mu_r\right),-\frac{1}{\beta_2}\left(\alpha_2-\lambda_r\right) \widehat p_0^{(\xi_p, \xi_{q})}\left(\mu_r\right), s\widehat  q_1^{(\xi_p, \xi_{q})}\left(\mu_r\right), \right. \\
	&\qquad  \left.   -\frac{1}{\beta_2} s\left(\alpha_2-\lambda_r\right) \widehat p_1^{(\xi_p, \xi_{q})}\left(\mu_r\right),  \ldots, -\frac{1}{\beta_2} (s)^{m-1}\left(\alpha_2-\lambda_r\right) \widehat p_{m-1}^{(\xi_p, \xi_{q})}\left(\mu_r\right)\right )^{\top}
\end{aligned}
\]
with $\xi_{p}=C, \xi_{q}=C$ for a certain constant $C$. 

Moreover, except for a few $\lambda_{r}$'s,  in all the above cases, we have 
\begin{equation}\babs{\widehat q_{k}^{(\xi_p, \xi_{q})}(\mu_{r})}\lesssim k, \quad \babs{\widehat p_{k}^{(\xi_p, \xi_{q})}(\mu_{r})}\lesssim k.
\end{equation}
\end{thm}
\begin{proof} The proof is divided into four steps.

\noindent \textbf{Step 1.} We consider 
\begin{equation}\label{equ:proofvectorlocalizedmode1}
\left(C_{2m+1, 2m+1}-\lambda_{r}I\right)\vect x = 0.
\end{equation}
We separate $\vect x$ as 
\[
\vect x = \begin{pmatrix}
	\vect y_1\\
	\eta_1\\
	\eta_2\\
	\vect y_2
\end{pmatrix},
\]	
where $\vect y_1$ is of size $2m$, $\vect y_2$ is of size $2m$ and $\eta_1$ and $\eta_2$ are two constants. Considering the last $2m+1$ rows in (\ref{equ:proofvectorlocalizedmode1}), we first have 
\begin{equation}\label{equ:proofvectorlocalizedmode2}
\left(\begin{array}{cccccc}
	 \gamma_{2} & \alpha_{1}-\lambda_{r}  & \beta_{1} & & &\\
	 & \gamma_{1} & \alpha_{2}-\lambda_{r}  & \ddots & &\\
	 & & \ddots & \ddots&\ddots&\\
	 & &  &\gamma_{1} & \alpha_2-\lambda_{r}  & \beta_2\\
	 & &  & & \gamma_{2} & \alpha_{1}-\lambda_{r} \\
\end{array}\right)\begin{pmatrix}
\eta_1\\
\eta_{2}\\
\vect y_{2,1}\\
\vect y_{2,2}\\
\vdots\\
\vect y_{2, 2m}
\end{pmatrix}=0.
\end{equation}
Considering the first row in the above equation, we obtain that
\[
\gamma_{2}\eta_1 + (\alpha_{1}-\lambda_{r})\eta_2 
 + \beta_{1}\vect y_{2,1} =0. 
\]
If $\eta_2 \neq 0$, then this gives 
\[
\left(\alpha_{1}+ \frac{\gamma_{2}\eta_1}{\eta_{2}}-\lambda_{r}\right)\eta_{2} + \beta_{1}\vect y_{2,1}=0.
\]
Therefore, it follows that 
\[
\left(\begin{array}{cccccc}
 \alpha_{1}+ \frac{\gamma_{2}\eta_1}{\eta_{2}}-\lambda_{r}  & \beta_{1} & & &\\
	\gamma_{1} & \alpha_{2}-\lambda_{r}  & \ddots & &\\
	 & \ddots & \ddots&\ddots&\\
   &  &\gamma_{1} & \alpha_2-\lambda_{r}  & \beta_2\\
	 &  & & \gamma_{2} & \alpha_{1}-\lambda_{r} \\
\end{array}\right)
	\begin{pmatrix}
\eta_{2}\\
\vect y_{2,1}\\
\vect y_{2,2}\\
\vdots\\
\vect y_{2, 2m}
\end{pmatrix}
=0.
\]
By Theorem \ref{thm:eigenvectoroddmatrixtwoperturb2}, we thus have
\begin{equation*}
	\begin{aligned}
	&\left(\eta_{2},\ \vect y_{2,1},\ \vect y_{2,2},\ \cdots, \ \vect y_{2, 2n}\right)^{\top}	\\
  =  &\left( \widehat q_{0}^{(\xi_p, \xi_{q})}\left(\mu_r\right),-\frac{1}{\beta_1}\left(\alpha_1-\lambda_r\right) \widehat p_0^{(\xi_p, \xi_{q})}\left(\mu_r\right), s \widehat q_1^{(\xi_p, \xi_{q})}\left(\mu_r\right),  -\frac{1}{\beta_1} s\left(\alpha_1-\lambda_r\right) \widehat p_1^{(\xi_p, \xi_{q})}\left(\mu_r\right), \right. \\
		&\qquad  \left.  \ldots, -\frac{1}{\beta_1} (s)^{m-1}\left(\alpha_1-\lambda_r\right) \widehat p_{m-1}^{(\xi_p, \xi_{q})}\left(\mu_r\right), (s)^m \widehat q_m^{(\xi_p, \xi_{q})}\left(\mu_r\right)\right )^{\top},
	\end{aligned}.
\end{equation*}
where $ \xi_{q}=(\alpha_1 -\lambda_r)C, \xi_{p}=\left(\alpha_{1}+ \frac{\gamma_{2}\eta_1}{\eta_2}-\lambda_{r}\right)C$ for a certain constant $C$. Moreover, $\lambda_{r}$ should be an eigenvalue of the above matrix.  
For the case when $\eta_2=0$, considering the second rows to the last rows of (\ref{equ:proofvectorlocalizedmode2}), we get 
\[
\left(\begin{array}{cccccc}
	\alpha_{2}-\lambda_{r}  & \beta_{2} & & &\\
	\gamma_{2} & \alpha_{1}-\lambda_{r}  & \beta_1 & &\\
	& \ddots & \ddots&\ddots&\\
	&  &\gamma_{1} & \alpha_2-\lambda_{r}  & \beta_2\\
	&  & & \gamma_{2} & \alpha_{1}-\lambda_{r} \\
\end{array}\right)
\begin{pmatrix}
	\vect y_{2,1}\\
	\vdots\\
	\vect y_{2,2m}
	\end{pmatrix}
=0.
\]
By Theorem \ref{thm:eigenvectorevenmatrixtwoperturb2}, we have 
\[
\begin{aligned}
&\left(\vect y_{2,1}, \cdots, \vect y_{2,2n}\right)^{\top} \\
=& \left( \widehat q_{0}^{(\xi_p, \xi_{q})}\left(\mu_r\right),-\frac{1}{\beta_2}\left(\alpha_2-\lambda_r\right) \widehat p_0^{(\xi_p, \xi_{q})}\left(\mu_r\right), s \widehat q_1^{(\xi_p, \xi_{q})}\left(\mu_r\right),  -\frac{1}{\beta_2} s\left(\alpha_2-\lambda_r\right) \widehat p_1^{(\xi_p, \xi_{q})}\left(\mu_r\right), \right. \\
&\qquad  \left.  \ldots, -\frac{1}{\beta_2} (s)^{m-1}\left(\alpha_2-\lambda_r\right) \widehat p_{m-1}^{(\xi_p, \xi_{q})}\left(\mu_r\right)\right )^{\top},
\end{aligned}
\]
where $\xi_{p}=C, \xi_{q}=C$ for some constant $C$. Moreover, $\lambda_{r}$ should be an eigenvalue of the matrix above. 

\medskip
\noindent \textbf{Step 2.} Considering the first $2m+1$ rows in (\ref{equ:proofvectorlocalizedmode1}), we have 
\begin{equation}\label{equ:proofvectorlocalizedmode3}
	\left(\begin{array}{cccccc}
		 \alpha_{1}-\lambda_{r}  & \gamma_{2} & & & &\\
		\beta_{2} & \alpha_{2}-\lambda_{r}  & \ddots & &&\\
		 & \ddots & \ddots&\ddots&&\\
		 &  &\beta_{2} & \alpha_2-\lambda_{r}  & \gamma_1&\\
		 &  & & \beta_{1} & \alpha_{1}-\lambda_{r}& \gamma_{2}\\
	\end{array}\right)\begin{pmatrix}
		\vect y_{1,1}\\
		\vect y_{1,2}\\
		\vdots\\
		\vect y_{1, 2m}\\
		\eta_1\\
            \eta_2
	\end{pmatrix}=0.
\end{equation}
Considering the last row in the above equation, we have 
\[
\gamma_{2}\eta_2 + (\alpha_{1}-\lambda_{r})\eta_{1} + \beta_{1}\vect y_{1,2m} =0. 
\]
If $\eta_{1}\neq 0$, this gives 
\[
\left(\alpha_{1}+ \frac{\gamma_{2}\eta_2}{\eta_1}-\lambda_{r}\right)\eta_1 + \beta_{1}\vect y_{1,2m}=0,
\]
and therefore, 
\[
	\left(\begin{array}{ccccc}
	\alpha_{1}-\lambda_{r}  & \gamma_{2} & & & \\
	\beta_{2} & \alpha_{2}-\lambda_{r}  & \ddots & &\\
	& \ddots & \ddots&\ddots&\\
	&  &\beta_{2} & \alpha_2-\lambda_{r}  & \gamma_1\\
	&  & & \beta_{1} & \alpha_{1}+ \frac{\gamma_{2}\eta_2}{\eta_1}-\lambda_{r}\\
\end{array}\right)\begin{pmatrix}
\vect y_{1,1}\\
\vect y_{1,2}\\
\vdots\\
\vect y_{1, 2m}\\
\eta_1
\end{pmatrix}=0.
\]
Note that the above equation corresponds to $A_{2m+1}^{(0, \frac{\gamma_{2}\eta_2}{\eta_1})}(\alpha_{1}, \gamma_{2}, \beta_{2}, \alpha_{2}, \gamma_{1}, \beta_{1})$, which is 
\[
R_{2m+1} A_{2m+1}^{(\frac{\gamma_{2}\eta_2}{\eta_1},0)}(\alpha_{1}, \beta_{1}, \gamma_{1}, \alpha_{2}, \beta_{2}, \gamma_{2})R_{2m+1}.
\]
By Theorem \ref{thm:eigenvectoroddmatrixtwoperturb2}, the eigenvector of $A_{2m+1}^{(\frac{\gamma_{2}\eta_2}{\eta_1},0)}(\alpha_{1}, \beta_{1}, \gamma_{1}, \alpha_{2}, \beta_{2}, \gamma_{2})$ has the following form: 
\begin{align*}
&\left(\widehat q_{0}^{(\xi_p, \xi_{q})}\left(\mu_r\right),-\frac{1}{\beta_1}\left(\alpha_1-\lambda_r\right) \widehat p_0^{(\xi_p, \xi_{q})}\left(\mu_r\right), s \widehat q_1^{(\xi_p, \xi_{q})}\left(\mu_r\right),  -\frac{1}{\beta_1} s\left(\alpha_1-\lambda_r\right) \widehat p_1^{(\xi_p, \xi_{q})}\left(\mu_r\right), \ldots, \right. \\
&\qquad  \left.-\frac{1}{\beta_1} \left(s\right)^{m-1}\left(\alpha_1-\lambda_r\right) \widehat p_{m-1}^{(\xi_p, \xi_{q})}\left(\mu_r\right), \left(s\right)^m \widehat q_m^{(\xi_p, \xi_{q})}\left(\mu_r\right)\right)^{\top},
\end{align*}
where $ \xi_{q}=(\alpha_1 -\lambda_r)C, \xi_{p}=\left(\alpha_{1}+ \frac{\gamma_{2}\eta_2}{\eta_1}-\lambda_{r} \right)C$ for a constant $C$. Therefore, by (\ref{equ:eigenvectorreverseoddmatrixtwoperturb3}), we have 
\begin{align*}
&\left(\vect y_{1,1}, \vect y_{1,2}, \cdots, \vect y_{1, 2m}, \eta_1\right)^{\top}\\
=& \left( \left(s\right)^m \widehat q_m^{(\xi_p, \xi_{q})}\left(\mu_r\right), -\frac{1}{\beta_1} \left(s\right)^{m-1}\left(\alpha_1-\lambda_r\right) \widehat p_{m-1}^{(\xi_p, \xi_{q})}\left(\mu_r\right),\cdots,-\frac{1}{\beta_1} s\left(\alpha_1-\lambda_r\right) \widehat p_1^{(\xi_p, \xi_{q})}\left(\mu_r\right),  \right. \nonumber \\
&\left.  \qquad s \widehat q_1^{(\xi_p, \xi_{q})}\left(\mu_r\right), -\frac{1}{\beta_1}\left(\alpha_1-\lambda_r\right) \widehat p_0^{(\xi_p, \xi_{q})}\left(\mu_r\right), \widehat q_{0}^{(\xi_p, \xi_{q})}\left(\mu_r\right)\right)^{\top},
\end{align*}
where $\xi_{q}=(\alpha_1-\lambda_r)C, \xi_{p}=\left( \alpha_{1}+ \frac{\gamma_{2}\eta_2}{\eta_1}-\lambda_{r}  \right)C$ for a certain constant $C$.  Moreover, $\lambda_{r}$ should be an eigenvalue of the above matrix.  

\medskip
For the case when $\eta_1=0$, considering the first row to the $2m$ rows of (\ref{equ:proofvectorlocalizedmode3}), we get 
\[
	\left(\begin{array}{cccccc}
	\alpha_{1}-\lambda_{r}  & \gamma_{2} & & \\
	\beta_{2} & \alpha_{2}-\lambda_{r}  & \gamma_1 &\\
	& \ddots & \ddots&\ddots\\
	&  &\beta_{1} & \alpha_2-\lambda_{r} \\
\end{array}\right)
\begin{pmatrix}
	\vect y_{1,1}\\
	\vdots\\
	\vect y_{1,2m}
\end{pmatrix}
=0.
\]
The above equation corresponds to $A_{2m}^{(0,0)}(\alpha_1, \gamma_2, \beta_2, \alpha_2, \gamma_1, \beta_1)$ which is 
\[
R_{2m} A_{2m}^{(0,0)}(\alpha_{2}, \beta_{1}, \gamma_{1}, \alpha_{1}, \beta_{2}, \gamma_{2})R_{2m}.
\]
By Theorem \ref{thm:eigenvectorevenmatrixtwoperturb2}, the eigenvector of $A_{2m}^{(0,0)}(\alpha_{2}, \beta_{1}, \gamma_{1}, \alpha_{1}, \beta_{2}, \gamma_{2})$ has the following form: 
\begin{align*}
	&\left(\widehat q_{0}^{(\xi_p, \xi_{q})}\left(\mu_r\right),-\frac{1}{\beta_1}\left(\alpha_2-\lambda_r\right) \widehat p_0^{(\xi_p, \xi_{q})}\left(\mu_r\right), s \widehat q_1^{(\xi_p, \xi_{q})}\left(\mu_r\right),  -\frac{1}{\beta_1} s\left(\alpha_2-\lambda_r\right) \widehat p_1^{(\xi_p, \xi_{q})}\left(\mu_r\right),\right. \\
	&\qquad   \ldots,  \left.-\frac{1}{\beta_1} \left(s\right)^{m-1}\left(\alpha_2-\lambda_r\right) \widehat p_{m-1}^{(\xi_p, \xi_{q})}\left(\mu_r\right)\right)^{\top},
\end{align*}
where $ \xi_{q}=C, \xi_{p}=C$ for a certain constant $C$. Therefore, analogously to  (\ref{equ:eigenvectorreverseoddmatrixtwoperturb3}), we have 
\begin{align*}
	&\left(\vect y_{1,1}, \vect y_{1,2},\cdots, \vect y_{1,2m}\right)^{\top}\\
	 =& \left( -\frac{1}{\beta_1} \left(s\right)^{m-1}\left(\alpha_2-\lambda_r\right) \widehat p_{m-1}^{(\xi_p, \xi_{q})}\left(\mu_r\right),\cdots,-\frac{1}{\beta_1} s\left(\alpha_2-\lambda_r\right) \widehat p_1^{(\xi_p, \xi_{q})}\left(\mu_r\right),  \right. \nonumber \\
	&\left.  \qquad s \widehat q_1^{(\xi_p, \xi_{q})}\left(\mu_r\right), -\frac{1}{\beta_1}\left(\alpha_2-\lambda_r\right) \widehat p_0^{(\xi_p, \xi_{q})}\left(\mu_r\right), \widehat q_{0}^{(\xi_p, \xi_{q})}\left(\mu_r\right)\right)^{\top},
\end{align*}
where $\xi_{q}=C, \xi_{p}=C$ for a certain constant $C$.  Moreover, $\lambda_{r}$ should be an eigenvalue of the above matrix.

\textbf{Step 3.} The rest of the proof is to control $\mu_r$ in order to control the quantities $\widehat q_{k}^{(\xi_p, \xi_{q})}\left(\mu_r\right)$, $\widehat p_{k}^{(\xi_p, \xi_{q})}\left(\mu_r\right)$. Although by the above discussions, each $\lambda_{r}$ should be an eigenvalue of a tridiagonal $2$-Toeplitz matrix with perturbations on the diagonal corners, we cannot directly apply Theorems \ref{thm:eigenvaluethm1} and \ref{thm:eigenvaluethm2} to control the eigenvalues and the values of the polynomials, as the perturbations on the corners vary for different $\lambda_{r}$. Note that we only need to prove, except for a few $\lambda_{r}$'s, that
\begin{equation}\label{equ:proofvectorlocalizedmode4a}
\mu_{r} =\cos \theta_{r}, \quad \theta_r \in [0, \pi].
\end{equation}
The idea is to first consider the principal submatrix of $C_{2m+1, 2m+1}$, that is, 
\[
D_{2m, 2m} = \left(
\begin{array}{ll}
     {\widehat G}_{11} &  {\widehat G}_{12} \\
      {\widehat G}_{21}&  {\widehat G}_{22}  
\end{array}
\right),
\]
where $ {\widehat G}_{11}=R_{2m} A_{2m}^{(0,0)}R_{2m}$, ${\widehat G}_{22} = A_{2m}^{(0,0)}$, and  
\[
 {\widehat G}_{12}  = \begin{pmatrix}
	0&0&\cdots &0\\
	\vdots&\vdots&\vdots& \vdots\\
	0&0&\cdots &0\\
	\gamma_{2}&0&\cdots&0
\end{pmatrix}, \quad {\widehat G}_{21}  = \begin{pmatrix}
0&\cdots&0 &\gamma_{2}\\
\vdots&\vdots&\vdots& \vdots\\
0&\cdots &0&0\\
0&\cdots&0&0
\end{pmatrix},
\] 
and then to analyse its eigenvalues. In particular, expanding the determinant $\babs{xI - D_{2m, 2m}}$ of $xI - D_{2m, 2m}$
by the Laplace method of expansion of determinants, we have that
\begin{align*}
&\babs{xI - A_{2m}^{(0,0)}}\babs{xI- A_{2m}^{(0,0)}}\\
&\quad -\gamma_2^2\babs{xI- A_{2m-1}^{(0,0)}(\alpha_2, \gamma_1, \beta_1, \alpha_1, \gamma_2, \beta_2)}\babs{xI- A_{2m-1}^{(0,0)}(\alpha_2, \gamma_1, \beta_1, \alpha_1, \gamma_2, \beta_2)}.
\end{align*}
By (\ref{equ:eigenpolynomial1}) and (\ref{equ:eigenpolynomial2}), the above determinant can be written as 
\begin{align*}
\left(P_m^*\left(\pi_2(x)\right)+\gamma_{2}\beta_{2} P_{m-1}^*\left(\pi_2(x)\right)\right)^2 -\gamma_2^2\left(x-\alpha_{2}\right)^2 P_{m-1}^*\left(\pi_2(x)\right)^2,
\end{align*}
where $P_{k}^*$ is defined by (\ref{equ:defiofpkstar1}). In particular, to find the roots of the above polynomial, we only need to solve the equation
\begin{align*}
\left(\sqrt{\gamma_{1}\beta_1 \gamma_2\beta_{2}}U_m\left(y(x)\right) + \gamma_2\beta_{2}U_{m-1}\left(y(x)\right) \right)^2  - \gamma_2^2\left(x-\alpha_{2}\right)^2 U_{m-1}\left(y(x)\right)^2=0,
\end{align*}
where $y(x)$ is defined by (\ref{equ:normalizedfunction1}). It is clear that $x$ such that $U_{m-1}(y(x))=0$ is not the solution to the above equation. We consider solving 
\begin{align}
\left(\frac{U_m\left(y(x)\right)}{U_{m-1}(y(x))} + \sqrt{\frac{\gamma_2\beta_{2}}{\gamma_1\beta_1}} \right)^2 = \frac{\gamma_2^2\left(x-\alpha_{2}\right)^2}{\sqrt{\gamma_{1}\beta_1 \gamma_2\beta_{2}}},
\end{align}
which gives 
\begin{equation}\label{equ:proofvectorlocalizedmode4}
\frac{U_m\left(y(x)\right)}{U_{m-1}(y(x))} + \sqrt{\frac{\gamma_2\beta_{2}}{\gamma_1\beta_1}} = + \frac{\gamma_2\left(x-\alpha_{2}\right)}{\left(\gamma_{1}\beta_1 \gamma_2\beta_{2}\right)^{\frac{1}{4}}}
\end{equation}
or 
\begin{equation}\label{equ:proofvectorlocalizedmode5}
\frac{U_m\left(y(x)\right)}{U_{m-1}(y(x))} + \sqrt{\frac{\gamma_2\beta_{2}}{\gamma_1\beta_1}} = - \frac{\gamma_2\left(x-\alpha_{2}\right)}{\left(\gamma_{1}\beta_1 \gamma_2\beta_{2}\right)^{\frac{1}{4}}}.
\end{equation}
In the following, we analyse the solutions to (\ref{equ:proofvectorlocalizedmode4}). By (\ref{equ:normalizedfunction1}), we can write
\[
x = g^+(y) : =  \frac{(\alpha_{1}+\alpha_{2})+\sqrt{(\alpha_{1}+\alpha_{2})^2+4((\gamma_{1}\beta_1+\gamma_{2}\beta_2-\alpha_1\alpha_2)+2y \sqrt{\gamma_{1}\beta_1 \gamma_{2}\beta_2} )}}{2}
\]
and 
\[
x = g^{-}(y) : =  \frac{(\alpha_{1}+\alpha_{2})-\sqrt{(\alpha_{1}+\alpha_{2})^2+4((\gamma_{1}\beta_1+\gamma_{2}\beta_2-\alpha_1\alpha_2)+2y \sqrt{\gamma_{1}\beta_1 \gamma_{2}\beta_2} )}}{2}.
\]
Now, (\ref{equ:proofvectorlocalizedmode4}) becomes
\begin{equation}\label{equ:proofvectorlocalizedmode6}
\left(\frac{U_m\left(y\right)}{U_{m-1}(y)} + \sqrt{\frac{\gamma_2\beta_{2}}{\gamma_1\beta_1}} \right)  = \frac{\gamma_2\left(g^{+}(y)-\alpha_{2}\right)}{\left(\gamma_{1}\beta_1 \gamma_2\beta_{2}\right)^{\frac{1}{4}}}
\end{equation}
and 
\begin{equation}\label{equ:proofvectorlocalizedmode7}
\left(\frac{U_m\left(y\right)}{U_{m-1}(y)} + \sqrt{\frac{\gamma_2\beta_{2}}{\gamma_1\beta_1}} \right) = \frac{\gamma_2\left(g^{-}(y)-\alpha_{2}\right)}{\left(\gamma_{1}\beta_1 \gamma_2\beta_{2}\right)^{\frac{1}{4}}}.
\end{equation}
We also suppose that $m$ is odd. The case when $m$ is even can be handled in a similar way, as discussed below. Note that the roots of $U_{m-1}(y)$ are $\cos \frac{k\pi}{m}, \ k=1,2,\cdots, m-1$ and 
\begin{equation*}
\begin{array}{ll}
U_{m-1}(y)> 0, &\text{for $y \in \left(\cos \left(\frac{(2k+1)\pi}{m}\right),\ \cos \left(\frac{2k\pi}{m}\right)\right), \ k=0, \cdots, \frac{m-1}{2},$} \\
U_{m-1}(y)< 0, &\text{for $y \in \left(\cos \left(\frac{(2k+2)\pi}{m}\right),\ \cos \left(\frac{(2k+1)\pi}{m}\right) \right),  \ k=0, \cdots, \frac{m-3}{2}.$}
\end{array}
\end{equation*}
The roots of $U_{m}(y)$ are $\cos \frac{k\pi}{m+1}, \ k=1,2,\cdots, m$ and 
\begin{equation*}
\begin{array}{ll}
U_{m}(y)> 0, &\text{for $y \in \left(\cos \left(\frac{(2k+1)\pi}{m+1}\right),\ \cos \left(\frac{2k\pi}{m+1}\right)\right), \ k=0, \cdots, \frac{m-1}{2},$} \\
U_{m}(y)< 0, &\text{for $y \in \left(\cos \left(\frac{(2k+2)\pi}{m+1}\right),\ \cos \left(\frac{(2k+1)\pi}{m+1}\right) \right),  \ k=0, \cdots, \frac{m-1}{2}.$}
\end{array}
\end{equation*}
Note also that
\[
 \frac{k}{m+1}< \frac{k}{m}< \frac{k+1}{m+1}, \quad k=1,\cdots,m-1,
\]
and  
\[
\cos\left(\frac{(k+1)\pi}{m+1}\right) < \cos\left(\frac{k\pi}{m}\right) < \cos\left(\frac{k\pi}{m+1}\right), \quad k=1,\cdots,m-1.
\]
 Thus, we have 
\begin{align*}
\lim_{y \rightarrow \cos\left(\frac{(m-1)\pi}{m}\right)^{-}}\frac{U_{m}(y)}{U_{m-1}(y)}=-\infty,&& \lim_{y \rightarrow \cos\left(\frac{(m-1)\pi}{m}\right)^{+}}\frac{U_{m}(y)}{U_{m-1}(y)}=+\infty, \\
\lim_{y \rightarrow \cos\left(\frac{(m-2)\pi}{m}\right)^{-}}\frac{U_{m}(y)}{U_{m-1}(y)}=-\infty,&& \lim_{y \rightarrow \cos\left(\frac{(m-2)\pi}{m}\right)^{+}}\frac{U_{m}(y)}{U_{m-1}(y)}=+\infty, \\
&\qquad\vdots&\\
\lim_{y \rightarrow \cos\left(\frac{\pi}{m}\right)^{-}}\frac{U_{m}(y)}{U_{m-1}(y)}=-\infty,&& \lim_{y \rightarrow \cos\left(\frac{\pi}{m}\right)^{+}}\frac{U_{m}(y)}{U_{m-1}(y)}=+\infty.
\end{align*}
Therefore, (\ref{equ:proofvectorlocalizedmode6}) must have at least one solution in each $\left(\cos\left(\frac{(k+1)\pi}{m}\right), \cos\left(\frac{k\pi}{m}\right)\right), \ k=1, \cdots, m-2$ and so does (\ref{equ:proofvectorlocalizedmode7}).
For $m$ even, we can prove the desired result in the same way. Thus we can assert that there are at least $2m-4$ solutions $\widehat \lambda_r$ of (\ref{equ:proofvectorlocalizedmode4}) satisfying $y(\widehat \lambda_r)=\cos \theta_r$. In the same manner, we can prove there are at least $2m-4$ solutions $\widehat \lambda_r$ of (\ref{equ:proofvectorlocalizedmode5}) satisfying that $y(\widehat \lambda_r)=\cos \theta_r$. Therefore, for at least $4m-8$ eigenvalues of $D_{2m, 2m}$, we have
\[
\babs{y(\widehat \lambda_j)}\leq 1, \quad j=1,\cdots, 4m-8
\]
with $y(x)$ being defined by (\ref{equ:normalizedfunction1}). 

\textbf{Step 4.} Furthermore, by the Cauchy interlacing theorem, we can show that, except for a few $\lambda_r$'s of $C_{2m+1, 2m+1}$, we have $\mu_r = y(\lambda_r) =\cos\theta_r$. Then by the same arguments as those in the proof of Theorem \ref{thm:skineffect1}, we can show that, for except a few $\lambda_{r}$'s, we have $\babs{\widehat q_{k}(\mu_{r})}\lesssim k, \babs{\widehat p_{k}(\mu_{r})}\lesssim k$. This completes the proof. 
\end{proof}

\section{Non-Hermitian skin effect and localised interface modes} \label{sect5}
Recently, a mathematical model of the skin effect in non-Hermitian chains of subwavelength resonators with an imaginary gauge potential has been developed \cite{ammari2023mathematical}. The authors were able to develop a rigorous theory for systems composed by equally spaced identical subwavelength resonators using the simple structure of
the gauge capacitance matrix associated with such structures and the rich literature on (tridiagonal) Toeplitz matrices and perturbations thereof.

In this section, we briefly recall the setup and established new results on dimer systems showing that the theory developed in the previous sections can be applied to study them. In particular, we prove condensation of the eigenmodes at one edge of the structure, find the topological origin of this phenomena and show that an 
interface formed by adjoining two half-structures with
opposite signs of complex gauge potentials leads to wave localisation along the interface. 

\subsection{Problem formulation}

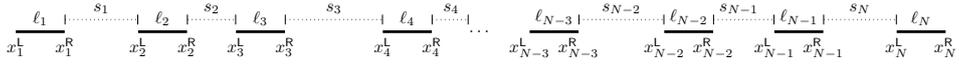
\begin{figure}[htb]
    \centering
    \begin{adjustbox}{width=\textwidth}
    \begin{tikzpicture}
        \coordinate (x1l) at (1,0);
        \path (x1l) +(1,0) coordinate (x1r);
        \path (x1r) +(0.75,0.7) coordinate (s1);
        \path (x1r) +(1.5,0) coordinate (x2l);
        \path (x2l) +(1,0) coordinate (x2r);
        \path (x2r) +(0.5,0.7) coordinate (s2);
        \path (x2r) +(1,0) coordinate (x3l);
        \path (x3l) +(1,0) coordinate (x3r);
        \path (x3r) +(1,0.7) coordinate (s3);
        \path (x3r) +(2,0) coordinate (x4l);
        \path (x4l) +(1,0) coordinate (x4r);
        \path (x4r) +(0.4,0.7) coordinate (s4);
        \path (x4r) +(1,0) coordinate (dots);
        \path (dots) +(1,0) coordinate (x5l);
        \path (x5l) +(1,0) coordinate (x5r);
        \path (x5r) +(1.75,0) coordinate (x6l);
        \path (x5r) +(0.875,0.7) coordinate (s5);
        \path (x6l) +(1,0) coordinate (x6r);
        \path (x6r) +(1.25,0) coordinate (x7l);
        \path (x6r) +(0.525,0.7) coordinate (s6);
        \path (x7l) +(1,0) coordinate (x7r);
        \path (x7r) +(1.5,0) coordinate (x8l);
        \path (x7r) +(0.75,0.7) coordinate (s7);
        \path (x8l) +(1,0) coordinate (x8r);
        \draw[ultra thick] (x1l) -- (x1r);
        \node[anchor=north] (label1) at (x1l) {$x_1^{\iL}$};
        \node[anchor=north] (label1) at (x1r) {$x_1^{\iR}$};
        \node[anchor=south] (label1) at ($(x1l)!0.5!(x1r)$) {$\ell_1$};
        \draw[dotted,|-|] ($(x1r)+(0,0.25)$) -- ($(x2l)+(0,0.25)$);
        \draw[ultra thick] (x2l) -- (x2r);
        \node[anchor=north] (label1) at (x2l) {$x_2^{\iL}$};
        \node[anchor=north] (label1) at (x2r) {$x_2^{\iR}$};
        \node[anchor=south] (label1) at ($(x2l)!0.5!(x2r)$) {$\ell_2$};
        \draw[dotted,|-|] ($(x2r)+(0,0.25)$) -- ($(x3l)+(0,0.25)$);
        \draw[ultra thick] (x3l) -- (x3r);
        \node[anchor=north] (label1) at (x3l) {$x_3^{\iL}$};
        \node[anchor=north] (label1) at (x3r) {$x_3^{\iR}$};
        \node[anchor=south] (label1) at ($(x3l)!0.5!(x3r)$) {$\ell_3$};
        \draw[dotted,|-|] ($(x3r)+(0,0.25)$) -- ($(x4l)+(0,0.25)$);
        \node (dots) at (dots) {\dots};
        \draw[ultra thick] (x4l) -- (x4r);
        \node[anchor=north] (label1) at (x4l) {$x_4^{\iL}$};
        \node[anchor=north] (label1) at (x4r) {$x_4^{\iR}$};
        \node[anchor=south] (label1) at ($(x4l)!0.5!(x4r)$) {$\ell_4$};
        \draw[dotted,|-|] ($(x4r)+(0,0.25)$) -- ($(dots)+(-.25,0.25)$);
        \draw[ultra thick] (x5l) -- (x5r);
        \node[anchor=north] (label1) at (x5l) {$x_{N-3}^{\iL}$};
        \node[anchor=north] (label1) at (x5r) {$x_{N-3}^{\iR}$};
        \node[anchor=south] (label1) at ($(x5l)!0.5!(x5r)$) {$\ell_{N-3}$};
        \draw[dotted,|-|] ($(x5r)+(0,0.25)$) -- ($(x6l)+(0,0.25)$);
        \draw[ultra thick] (x6l) -- (x6r);
        \node[anchor=north] (label1) at (x6l) {$x_{N-2}^{\iL}$};
        \node[anchor=north] (label1) at (x6r) {$x_{N-2}^{\iR}$};
        \node[anchor=south] (label1) at ($(x6l)!0.5!(x6r)$) {$\ell_{N-2}$};
        \draw[dotted,|-|] ($(x6r)+(0,0.25)$) -- ($(x7l)+(0,0.25)$);
        \draw[ultra thick] (x7l) -- (x7r);
        \node[anchor=north] (label1) at (x7l) {$x_{N-1}^{\iL}$};
        \node[anchor=north] (label1) at (x7r) {$x_{N-1}^{\iR}$};
        \node[anchor=south] (label1) at ($(x7l)!0.5!(x7r)$) {$\ell_{N-1}$};
        \draw[dotted,|-|] ($(x7r)+(0,0.25)$) -- ($(x8l)+(0,0.25)$);
        \draw[ultra thick] (x8l) -- (x8r);
        \node[anchor=north] (label1) at (x8l) {$x_{N}^{\iL}$};
        \node[anchor=north] (label1) at (x8r) {$x_{N}^{\iR}$};
        \node[anchor=south] (label1) at ($(x8l)!0.5!(x8r)$) {$\ell_N$};
        \node[anchor=north] (label1) at (s1) {$s_1$};
        \node[anchor=north] (label1) at (s2) {$s_2$};
        \node[anchor=north] (label1) at (s3) {$s_3$};
        \node[anchor=north] (label1) at (s4) {$s_4$};
        \node[anchor=north] (label1) at (s5) {$s_{N-2}$};
        \node[anchor=north] (label1) at (s6) {$s_{N-1}$};
        \node[anchor=north] (label1) at (s7) {$s_N$};
    \end{tikzpicture}
    \end{adjustbox}
    \caption{A chain of $N$ subwavelength resonators, with lengths
    $(\ell_i)_{1\leq i\leq N}$ and spacings $(s_{i})_{1\leq i\leq N-1}$.}
    \label{fig:setting}
\end{figure}

We consider a one-dimensional chain of $N$ disjoint subwavelength resonators $D_i\coloneqq (x_i^{\iL},x_i^{\iR})\subset \R$, where $(x_i^{\iLR})_{1\leq i\leq N} \subset \R$ are the $2N$ extremities satisfying $x_i^{\iL} < x_i^{\iR} <  x_{i+1}^{\iL}$ for any $1\leq i \leq N$. We fix the coordinates such that $x_1^{\iL}=0$. We also denote by  $\ell_i = x_i^{\iR} - x_i^{\iL}$ the length of the $i$-th resonators,  and by $s_i= x_{i+1}^{\iL} -x_i^{\iR}$ the spacing between the $i$-th and $(i+1)$-th resonators. The system is illustrated in \cref{fig:setting}. 

We will use the notation $\displaystyle  D\coloneqq \bigcup_{i=1}^N(x_i^{\iL},x_i^{\iR})\subset \R$ to symbolise the set of subwavelength resonators and denote for a function $w$ 
$$ w \vert_{\iR}(x) = \lim_{s\rightarrow 0^+} w(x + s), \quad  w \vert_{\iL}(x) = \lim_{s\rightarrow 0^+} w(x - s).$$

We consider the following system of ordinary differential equations:
\begin{align}
    \begin{dcases}
        u\prii(x) + \gamma u\pri(x)+\frac{\omega^2}{v_b^2}u=0, & x\in D,\\
        u\prii(x) + \frac{\omega^2}{v^2}u=0, & x\in\R\setminus D,\\
        u\vert_{\iR}(x^{\iLR}_i) - u\vert_{\iL}(x^{\iLR}_i) = 0, & \text{for all } 1\leq i\leq N,\\
        \left.\frac{\dd u}{\dd x}\right\vert_{\iR}(x^{\iL}_{{i}})=\delta\left.\frac{\dd u}{\dd x}\right\vert_{\iL}(x^{\iL}_{{i}}), & \text{for all } 1\leq i\leq N,\\
        \left.\frac{\dd u}{\dd x}\right\vert_{\iR}(x^{\iR}_{{i}})=\delta\left.\frac{\dd u}{\dd x}\right\vert_{\iR}(x^{\iL}_{{i}}), & \text{for all } 1\leq i\leq N,\\
        \frac{\dd u}{\dd\vert x\vert } - \mathbf{i} \frac{\omega}{v} u = 0, & x\in(-\infty,x_1^{\iL})\cup (x_N^{\iR},\infty).
    \end{dcases}
\label{eq:coupled ods}
\end{align}
Here, the parameter $\gamma \in \mathbb{R}, \gamma \neq 0, $ models the complex gauge potential, $0<\delta \ll 1$ is a small contrast material parameter, the positive constants $v$ and $v_b$
are respectively the wave speeds outside and inside the resonators and $\omega$ denotes the frequency. We refer the reader to \cite[Section 2]{ammari2023mathematical} for the physical motivation of this model.

We are interested in the resonances $\omega\in \mathbb{C}$ such that \eqref{eq:coupled ods} has a non-trivial solution $u$. We look for the modes within the subwavelength regime, which we characterise by imposing $\omega \to 0$ as $\delta\to 0$. This regime will recover subwavelength resonances, while keeping the size of the resonators fixed.

A central result of \cite{ammari2023mathematical} is the approximation of the eigenfrequencies and eigenmodes of \eqref{eq:coupled ods} in the subwavelength regime with a finite dimensional eigenvalue problem as we recall in \cref{prop: approx via eva eve}. The matrix $ \mathcal{C}^\gamma  =\bigr(\mathcal{C}_{i,j}^\gamma \bigr)_{i,j}$ involved in such finite-dimensional approximation of \eqref{eq:coupled ods} is the so-called \emph{gauge capacitance matrix} and is given by
\begin{align}
    \mathcal{C}_{i,j}^\gamma \coloneqq \begin{dcases}
        \frac{\gamma}{s_1} \frac{\ell_1}{1-e^{-\gamma \ell_1}}, & i=j=1,\\
         \frac{\gamma}{s_i} \frac{\ell_i}{1-e^{-\gamma \ell_i}} -\frac{\gamma}{s_{i-1}} \frac{\ell_i}{1-e^{\gamma \ell_i}},  & 1< i=j< N,\\
       - \frac{\gamma}{s_i} \frac{\ell_i}{1-e^{-\gamma \ell_j}},  & 1\leq i=j-1\leq N-1,\\
       \frac{\gamma}{s_j} \frac{\ell_i}{1-e^{\gamma \ell_j}}, & 2\leq i=j+1\leq N,\\
      - \frac{\gamma}{s_{N-1}} \frac{\ell_N}{1-e^{\gamma \ell_N}}, & i=j=N,\\
    \end{dcases}\label{eq: explicit coef cap mat}
\end{align} 
while all the other entries are zero. The following result is from \cite[Corollary 2.6]{ammari2023mathematical}.
\begin{prop}[Discrete approximations of the eigenfrequencies and eigenmodes]\label{prop: approx via eva eve}
    The $N$ subwavelength eigenfrequencies $\omega_i$ satisfy, as $\delta\to0$,
    \begin{align*}
        \omega_i =  v_b \sqrt{\delta\lambda_i} + \mathcal{O}(\delta),
    \end{align*}
    where $(\lambda_i)_{1\leq i\leq N}$ are the eigenvalues of the eigenvalue problem
\begin{equation}
\label{eq:eigevalue problem capacitance matrix}
\mathcal{C}^\gamma \vect a_i = \lambda_i V \vect a_i,\qquad 1\leq i\leq N,
\end{equation}
where $V=\diag(\ell_1,\dots,\ell_N)$. Furthermore, let $u_i$ be a subwavelength eigenmode corresponding to $\omega_i$ and let $\vect a_i$ be the corresponding eigenvector of $\mathcal{C}^\gamma$. Then
        \begin{align*}
            u_i(x) = \sum_j \vect a_i^{(j)}V_j(x) + \mathcal{O}(\delta),
        \end{align*}
        where $V_j$ are the solutions of 
\begin{align}
    \begin{dcases}
        -\frac{\dd{^2}}{\dd x^2} V_j =0, & x\in\R\setminus D, \\
        V_i(x)=\delta_{ij}, & x\in (x_j^{\iL},x_j^{\iR}),\\
        V_i(x) = \mathcal{O}(1) & \mbox{as } \vert x\vert\to\infty,
    \end{dcases}
    \label{eq: def V_i}
\end{align}
with $\delta_{ij}$ being the Kronecker symbol and $\vect a^{(j)}$ denotes the $j$-th entry of the eigenvector.
\end{prop}
\begin{remark}
    Since $V_j$ is piecewise linear, supported in $(x_{j-1}^{\iR},x_{j+1}^{\iL})$ and $V_j(x)=1$ for $x\in(x_{j}^{\iL},x_{j}^{\iR})$ the overall behaviour of the eigenmodes $u_i$ is captured by the eigenvectors $\vect a_i$.
\end{remark}
\subsection{Non-Hermitian skin effect in dimer systems and condensation of the system's eigenmodes}\label{sec: condensation dimers}
For a system of dimers --- that is $s_i=s_{i+2}$ for all $1\leq i\leq N-3$ and $\ell_i=\ell$ for all $1\leq i\leq N$ --- the gauge capacitance matrix from \eqref{eq: explicit coef cap mat} takes the form
\begin{align}
    \mathcal{C}^\gamma =
    \begin{pmatrix}
    \tilde{\alpha}_{1} & \beta_{1} & & & & &\\
		\eta_{1} & \alpha_{2} & \beta_{2} & & & &\\
		& \eta_{2} & \alpha_{1} & \beta_{1} & & &\\
		& & \eta_{1} & \alpha_{2} & \ddots & &\\
		& & & \ddots & \ddots&\ddots&\\
		& & &  &\eta_{2} & \alpha_1 & \beta_1\\
		& & &  & & \eta_{1} & \tilde{\alpha}_{2}\\
    \end{pmatrix},
\end{align}
where
\begin{align*}
    \alpha_1&=\frac{\gamma}{s_1} \frac{\ell}{1-e^{-\gamma \ell}} -\frac{\gamma}{s_2} \frac{\ell}{1-e^{\gamma \ell}}, & \alpha_2&=\frac{\gamma}{s_2} \frac{\ell}{1-e^{-\gamma \ell}} -\frac{\gamma}{s_1} \frac{\ell}{1-e^{\gamma \ell}},\\
    \tilde{\alpha}_1&=\frac{\gamma}{s_1} \frac{\ell}{1-e^{-\gamma \ell}},& \tilde{\alpha}_2&=- \tilde{\alpha}_1,\\
    \beta_1&= - \frac{\gamma}{s_1} \frac{\ell}{1-e^{-\gamma \ell}},& \beta_2&=- \frac{\gamma}{s_2} \frac{\ell}{1-e^{-\gamma \ell}},\\
    \eta_1&=\frac{\gamma}{s_1} \frac{\ell }{1-e^{\gamma \ell }}, & \eta_2&=\frac{\gamma}{s_2} \frac{\ell }{1-e^{\gamma \ell }}.\\
\end{align*}

We use $\eta_i$ instead of $\gamma_i$ (as done in the previous sections) to denote the coefficients on the lower diagonal in order to prevent confusion. One may remark that all rows of $\mathcal{C}^\gamma$ sum to 0 and thus $\bm 1\in\ker(\mathcal{C}^\gamma)$. The next theorem states that $\bm 1$ is the only eigenvector of $\mathcal{C}^\gamma$ that is not localised.

\begin{thm}\label{thm: condensation of eigenmodes for dimer systems}
    All but a few (independent of $N$) eigenvectors of the gauge capacitance matrix $\mathcal{C}^\gamma$ satisfy the following inequality:
    \begin{equation}\label{eq: estimate eigevector cap mat}
\babs{\vect x^{(j)}}\leq Mj e^{-\ell\gamma\lfloor\frac{j-1}{2}\rfloor},
\end{equation}
where $\vect x^{(j)}$ is the $j$-th component of the eigenvector $\vect x$.
\end{thm}
Theorem \ref{thm: condensation of eigenmodes for dimer systems} is a direct consequence of Theorem \ref{thm:eigenvectorevenmatrixtwoperturb2}. 

\begin{figure}[h]
    \centering
    \begin{subfigure}[t]{0.48\textwidth}
        \includegraphics[width=1\textwidth]{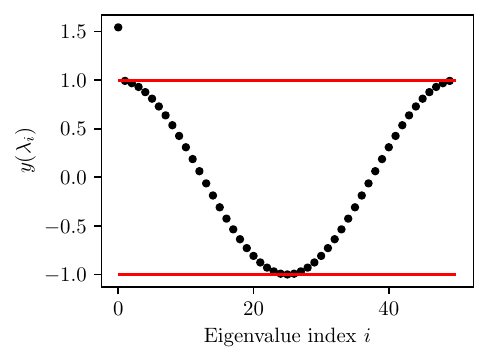}
    \caption{The black dots show the value $y(\lambda_i)$ for the eigenvalues $\lambda_i$ of $\mathcal{C}^\gamma$. The red line show the boundaries stability zone $y=\pm 1$. Only for $\lambda=0$, $y(\lambda)$ lays outside of this zone while Theorem \ref{thm:eigenvectorevenmatrixtwoperturb2} predicts that at most $10$ do.}
    \label{fig: criteria decay}
    \end{subfigure}\hfill
    \begin{subfigure}[t]{0.48\textwidth}
        \includegraphics[width=1\textwidth]{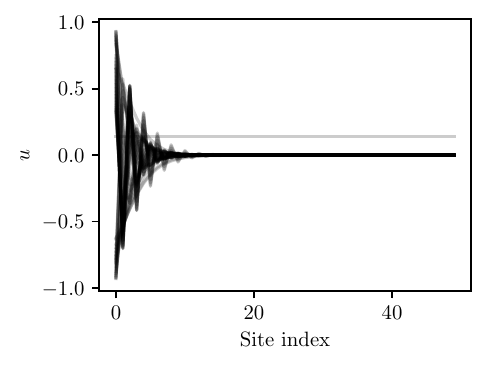}
    \caption{Eigenmodes superimposed on one another to portray the skin effect: all but one modes are exponentially localised on the left edge of the system.}
    \label{fig: skin effect dimers}
    \end{subfigure}
    \caption{Eigenmode localisation for a system of $25$ dimers ($N=50$), $\ell_i=1$, $s_1=1$, $s_2=2$ and $\gamma=1$.}
\end{figure}

Numerically, we can verify that there is  exactly one eigenvector which is not localised. More precisely, we show in Figure \ref{fig: criteria decay} that only the eigenvalue $\lambda_1=0$ satisfies $\vert y(\lambda_i)\vert \geq 1$ for $y$ as in \eqref{equ:normalizedfunction1}. In Figure \ref{fig: skin effect dimers}, we show the eigenmodes of a system of $25$ dimers.

It is interesting to remark that the eigenvalue $0$ is both the only outlier of Figure \ref{fig: criteria decay} and the only point laying on the trace of the eigenvalues of the symbol of the 2-Toeplitz operator in Figures \ref{fig: eig winding 50} and \ref{fig: eig winding 100}.

\subsection{Topological origin of the non-Hermitian skin effect in dimer systems}

\begin{figure}[p]
    \centering
    \begin{subfigure}[t]{0.48\textwidth}
        \centering
        \includegraphics[width=\textwidth]{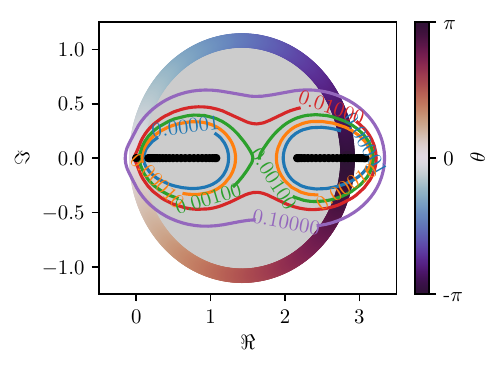}
        \caption{The winding of the determinant does not predict the exponential decay correctly. Simulation performed with a system of $25$ dimers.}
        \label{fig: det winding}
    \end{subfigure}
    \hfill
    \begin{subfigure}[t]{0.48\textwidth}
        \centering
        \includegraphics[width=\textwidth]{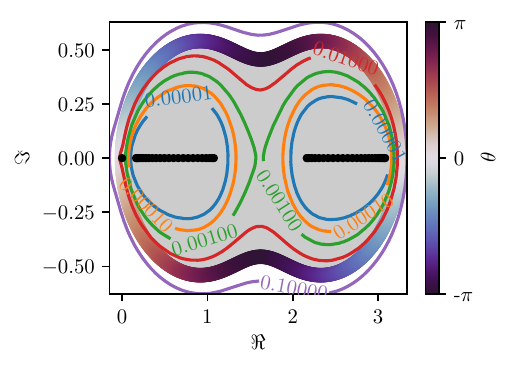}
        \caption{The winding of the eigenvalues does predict the exponential decay correctly. Simulation performed with a system of $25$ dimers.}
        \label{fig: eig winding 50}
    \end{subfigure}\\[5mm]
    \begin{subfigure}[t]{0.8\textwidth}
        \centering
        \includegraphics[width=\textwidth]{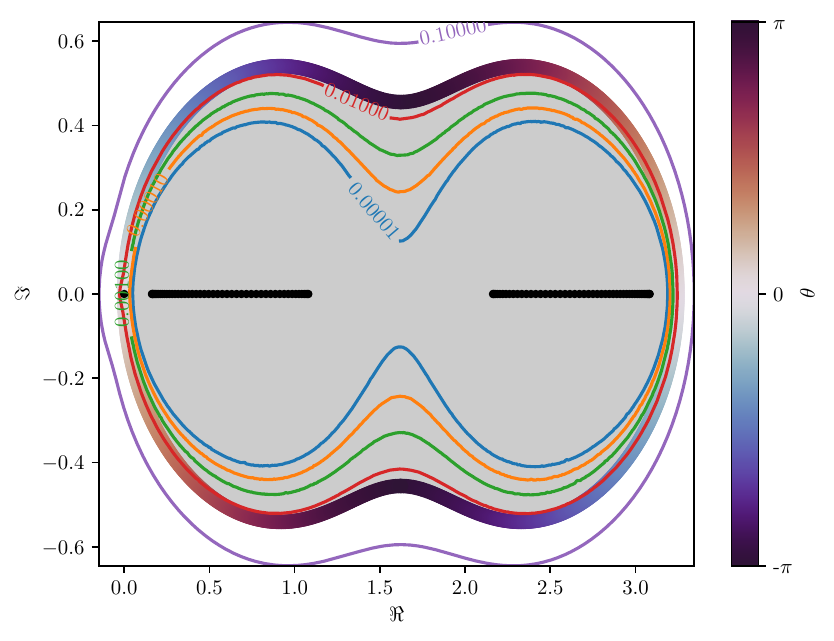}
        \caption{The region of non-trivial winding of the eigenvalues is the limit, as $N\to\infty$, of the $\varepsilon$-pseudospectrum for any fixed $\varepsilon$. Simulation performed with a system of $50$ dimers.}
        \label{fig: eig winding 100}
    \end{subfigure}
    \caption{Spectra and pseudospectra for dimer systems. In all three figures the black dots show the spectrum of $\mathcal{C}^\gamma$ and the solid colored lines the $\varepsilon$-pseudospectrum for $\varepsilon=10^k$ for $k=-1,\dots,-5$. In Figure \ref{fig: det winding}, the graduated line shows the winding of the determinant and in Figures \ref{fig: eig winding 50} and \ref{fig: eig winding 100}, the winding of the union of the eigenvalues of the symbol of the corresponding $2$-Toeplitz operator. In light grey, the respective region with non-trivial winding. All simulations are performed with $\ell_i=1$, $s_1=1$, $s_2=2$ and $\gamma=1$.}
    \label{fig: 3 figs spectra and winding}
\end{figure}

In the case of a perturbed Toeplitz matrix --- as for equally spaced and identical resonators --- it is well known that the exponential decay of the eigenvectors is due to the winding of the symbol of the corresponding Toeplitz operator (or equivalently, its Fredholm index) \cite{ammari2023mathematical, trefethen2005spectra}. To the best of our knowledge, no such result is known for $K$-Toeplitz matrices when $K \geq 2$. However, it is known (see, for instance, \cite{bottcher}) that the Fredholm index of the associated operator is given by the winding of the determinant of its symbol, which in this case is a $K\times K$ matrix, provided that the determinant does not vanish at any point on the unit circle. In Figure \ref{fig: det winding}, we show the spectrum and pseudospectrum of the gauge capacitance matrix of a system of $25$ dimers together with the trace of the determinant of the symbol of the associated $2$-Toeplitz operator in the complex plane. We observe that various eigenvalues whose eigenvectors are localised lay in a region without winding. This is due to the fact that the determinant of the symbol takes value zero at some point on the unit circle. 

In Figures \ref{fig: eig winding 50} and \ref{fig: eig winding 100}, we consider systems of $25$ and $50$ dimers, respectively. This time, the colored trace shows the winding of the two eigenvalues of the symbol of the corresponding $2$-Toeplitz operator. This winding predicts accurately the exponential decay of the eigenmodes and is the limit of the pseudospectrum as $N\to\infty$ as in the simplest case studied in \cite{ammari2023mathematical}.

\subsection{Non-Hermitian interface modes in systems with opposing signs of $\gamma$.}

We consider a system modelled by
\begin{align}
    \begin{dcases}
        u\prii(x) + \gamma_i u\pri(x)+\frac{\omega^2}{v_b^2}u=0, & x\in  (x_i^{\iL}, x_i^{\iR}),\\
        u\prii(x) + \frac{\omega^2}{v^2}u=0, & x\in\R\setminus D,\\
        u\vert_{\iR}(x^{\iLR}_i) - u\vert_{\iL}(x^{\iLR}_i) = 0, & \text{for all } 1\leq i\leq N,\\
        \left.\frac{\dd u}{\dd x}\right\vert_{\iR}(x^{\iL}_{{i}})=\delta\left.\frac{\dd u}{\dd x}\right\vert_{\iL}(x^{\iL}_{{i}}), & \text{for all } 1\leq i\leq N,\\
        \left.\frac{\dd u}{\dd x}\right\vert_{\iR}(x^{\iR}_{{i}})=\delta\left.\frac{\dd u}{\dd x}\right\vert_{\iR}(x^{\iL}_{{i}}), & \text{for all } 1\leq i\leq N,\\
        \frac{\dd u}{\dd\vert x\vert } -\mathbf{i} \frac{\omega}{v} u = 0, & x\in(-\infty,x_1^{\iL})\cup (x_N^{\iR},\infty).
    \end{dcases}
\label{eq:coupled ods gamma_i}
\end{align}
From \cite{ammari2023mathematical},  we know that the subwavelength eigenfrequencies and eigenmodes of the above system can be approximated by the eigenvalues and eigenvectors of the following gauge capacitance matrix $(\mathcal{C}^\gamma)_{i,j}$ defined by
\begin{align}
    \mathcal{C}_{i,j}^\gamma \coloneqq \begin{dcases}
        \frac{\gamma_1}{s_1} \frac{\ell_1}{1-e^{-\gamma_i \ell_1}}, & i=j=1,\\
         \frac{\gamma_i}{s_i} \frac{\ell_i}{1-e^{-\gamma_i \ell_i}} -\frac{\gamma_i}{s_{i-1}} \frac{\ell_i}{1-e^{\gamma_i \ell_i}},  & 1< i=j< N,\\
       - \frac{\gamma_i}{s_i} \frac{\ell_i}{1-e^{-\gamma_i \ell_j}},  & 1\leq i=j-1\leq N-1,\\
       \frac{\gamma_i}{s_j} \frac{\ell_i}{1-e^{\gamma_i \ell_j}}, & 2\leq i=j+1\leq N,\\
      - \frac{\gamma_i}{s_{N-1}} \frac{\ell_N}{1-e^{\gamma_i \ell_N}}, & i=j=N,\\
    \end{dcases}\label{eq: explicit coef cap mat gamma_i}
\end{align} 
while all the other entries are zero. We are particularly interested in the case where
\begin{align}
    \gamma_i = \begin{dcases}
        -\gamma, &  \text{for } 1\leq i \leq m,\\
        \gamma, & \text{for } m+1 \leq i \leq N,
    \end{dcases}
    \label{eq: gamma_is}
\end{align}
for some $\gamma>0$ and $1<m<N$, where we typically choose $2m=N$ for $m$ even, creating an symmetric structure. The gauge capacitance matrix then becomes 
\begin{align}
    \mathcal{C}^\gamma =
    \begin{pmatrix}
    \tilde{\alpha}_{1} & \eta_{1} & & & & &\\
		 \beta_{1}& \alpha_{2} & \eta_{2} & & & &\\
		& \beta_{2} & \alpha_{1} & \eta_{1} & & &\\
		 & & \ddots & \ddots&\ddots&\\
  && & \beta_{1} & \alpha_{2} & \eta_{2} & &\\
    &&& & \eta_{2} & \alpha_{1} & \beta_{1} & &\\
          &&&&& \ddots & \ddots&\ddots&\\
      &&&&& & \eta_{2} & \alpha_{1} & \beta_{1}\\
		&&&&&&& \eta_{1} & \tilde{\alpha}_{2}\\
    \end{pmatrix}.
\end{align}
Using the results from Subsection \ref{subsect4}, it is now possible to show that all but a few eigenmodes are localised around the interface.

\begin{thm}
    Consider a system satisfying \eqref{eq:coupled ods gamma_i} for $\gamma_i$ satisfying \eqref{eq: gamma_is}. Then all but a few eigenmodes are exponentially localised around the interface at the site index $m$. More explicitly,  we have
    \begin{align*}
    \vert \vect x^{(j)} \vert \leq M \vert m - j\vert e^{-\gamma\ell\left\vert \frac{m-j}{2}\right\vert},
    \end{align*}
    for all $1\leq j\leq N$ but $m\neq j$, with $M$ being some positive constant.
\end{thm}

\begin{figure}[h]
    \centering
    \includegraphics[width=0.5\textwidth]{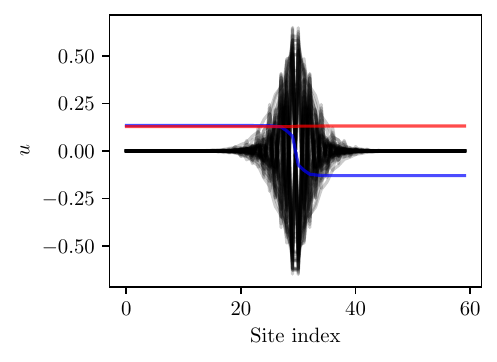}
    \caption{Interface modes for a system of $30$ resonators with negative $\gamma_i$ on the left and $30$ resonators with positive $\gamma_i$ on the right. Simulations performed with $\ell_i=1$, $s_1=1$, $s_2=2$ and $\vert\gamma_i\vert=1$.}
    \label{fig: interface modes}
\end{figure}

In Figure \ref{fig: interface modes}, we show that the interface modes for a system of $2\times 30$ resonators. One immediately recognises that all but two eigenmodes are exponentially localised around the interface. The non-localised eigenmodes present once a monopole and once a dipole behavior.

\section{Concluding remarks} \label{sect6}
Based on new explicit formulas for the eigenpairs of perturbed tridiagonal block Toeplitz matrices,  we  have analysed the non-Hermitian skin effect arising in  dimer systems of subwavelength resonators and shown its topological origin. We have also proved the localisation of interface modes between systems of resonators with imaginary gauge potentials with opposite signs.

This paper opens the door to the study of many-body non-Hermitian systems where the non-Hermiticity arises from complex gauge potentials.  The explicit theory we have developed for tridiagonal $2$-Toeplitz matrices could be extended to tridiagonal $K$-Toeplitz matrix and would lead to a generalisation of the obtained results to systems with arbitrary number of periodically repeated resonators. Another interesting problem is to estimate the stability of the non-Hermitian skin effect with respect to small changes in either the positions of the subwavelength resonators or their material parameters. This will be the subject of a forthcoming publication \cite{SkinStability}.

\addtocontents{toc}{\protect\setcounter{tocdepth}{0}}
\section*{Acknowledgments}
This work was partially supported by Swiss National Science Foundation grant number 200021--200307. 
\bigskip

\section*{Code availability}
The data that support the findings of this work are openly available at \\ \href{https://doi.org/10.5281/zenodo.8171204}{https://doi.org/10.5281/zenodo.8171204}.

\bibliographystyle{plain}
\bibliography{reference_final}	
\end{document}